\documentclass[aps,prl,reprint,superscriptaddress,preprintnumbers,amsmath,amssymb,dvipdfmx,
secnumarabic
]{revtex4-2}

\usepackage{comment} 
\usepackage[dvipdfmx]{graphicx}
\usepackage{dcolumn}
\usepackage{bm}
\usepackage{siunitx} 

\usepackage{textcomp}
\usepackage{mathcomp}
\usepackage{amsmath}

\usepackage{hyperref}
\RequirePackage{fix-cm}

\usepackage[mathlines]{lineno}
\usepackage{color}
\newcommand{\fixred}[1]{{\color{black}#1}}

\newcommand{\fixcyan}[1]{{\color{cyan}#1}}

\usepackage{epstopdf}

\begin{document}

\title{Arrested coarsening in active colloidal suspensions driven by nonreciprocal electrohydrodynamic interactions
}
	\author{Shoma Hara}
	\affiliation{%
		Department of Applied Physics, Tokyo University of Science, 6-3-1 Nijuku, Katsushika, Tokyo, 125-8585, Japan}
    \author{Masazumi Okada}
	\affiliation{%
		Department of Applied Physics, Tokyo University of Science, 6-3-1 Nijuku, Katsushika, Tokyo, 125-8585, Japan}
	\author{Keisuke Kittaka}
	\affiliation{%
		Department of Applied Physics, Tokyo University of Science, 6-3-1 Nijuku, Katsushika, Tokyo, 125-8585, Japan}
    \author{Sho Tanami}
	\affiliation{%
		Department of Applied Physics, Tokyo University of Science, 6-3-1 Nijuku, Katsushika, Tokyo, 125-8585, Japan}
	\author{Yuichi Iwasaki}
	\affiliation{%
		Department of Applied Physics, Tokyo University of Science, 6-3-1 Nijuku, Katsushika, Tokyo, 125-8585, Japan}
	\author{Hiroaki Ishikawa}
	\affiliation{%
		Department of Applied Physics, Tokyo University of Science, 6-3-1 Nijuku, Katsushika, Tokyo, 125-8585, Japan}
	\author{Kiwamu Yoshii}
	\affiliation{%
		Department of Applied Physics, Tokyo University of Science, 6-3-1 Nijuku, Katsushika, Tokyo, 125-8585, Japan}
	\author{Yutaka Sumino}
	\email{ysumino@rs.tus.ac.jp}
	\affiliation{%
		Department of Applied Physics, Tokyo University of Science, 6-3-1 Nijuku, Katsushika, Tokyo, 125-8585, Japan}
    \affiliation{Water Frontier Research Center and Division of Colloid Interface, Research Institute for Science \& Technology, Tokyo University of Science, 6-3-1 Nijuku, Katsushika, Tokyo, 125-8585, Japan
	}
    \affiliation{
 Faculty of Engineering and Physical Sciences, University of Surrey, Guildford, Surrey GU2 7XH, United Kingdom
}%

\date{\today}

\begin{abstract}
\fixred{
Nonreciprocal interactions have recently attracted growing interest in nonequilibrium physics. In particular, breaking action–reaction symmetry has been proposed as a mechanism for collective motion, yet controlled experimental realizations remain scarce. Here we show that bidisperse colloidal suspensions driven by AC electric fields exhibit persistent active clusters sustained by nonreciprocal electrohydrodynamic interactions. Size-asymmetric particle pairs spontaneously self-propel due to imbalanced electrohydrodynamic attraction, producing clusters that continuously fragment and reorganize rather than coarsening into static aggregates as in monodisperse systems. Agent-based simulations reproduce the observed dynamics and identify nonreciprocal pair propulsion as the minimal ingredient for the persistent cluster dynamics. These results demonstrate that action–reaction symmetry breaking in electrohydrodynamic interactions can arrest coarsening and sustain dynamically reconfigurable collective states in dense colloidal suspensions.
}
\end{abstract}

\maketitle

\setcounter{secnumdepth}{3}

Self-propelled inanimate particles~\cite{Paxton2004, Nagayama2004} have long attracted attention as elements that break momentum conservation and generate novel collective dynamics.
Such systems, broadly referred to as {\it active matter}~\cite{Marchetti2013, Menzel2015, Ramaswamy2010, Vicsek2012a}, extend statistical mechanics to far-from-equilibrium regimes and inspire micrometer-scale functional assemblies~\cite{Ginelli2016, Vicsek1995, Toner1995, Toner1998, Peshkov2014a, Kassem2017, Astumian2016, Wang2015b, Araki2021, Duclos2020, Keber2014-jm, Sanchez2012}.
Nonreciprocal interactions~\cite{Fruchart2021} have recently attracted considerable interest in nonequilibrium physics. 
In active matter, such interactions break action–reaction symmetry and generate behaviors inaccessible to reciprocal systems~\cite{Agudo-Canalejo2019}, 
\fixred{yet experimental realizations have so far been limited to small clusters ($N \le 5$)~\cite{Meredith2020-vz, Niu2017, Niu2018, Ishikawa2022-mx,Sumino2023-yr, Ishikawa2025-tf}, and remain scarce at collective scales ($N \geq 10^4$).
}
Here we show that bidisperse colloids under AC electric fields exhibit persistent active clusters sustained by nonreciprocal electrohydrodynamic (EHD) interactions~\cite{Ristenpart2003, Ristenpart2004, Ristenpart2007d}. When particles of different sizes are mixed, the EHD-induced flows generate asymmetric forces~\cite{Ma2015, Niu2017, Niu2018, Grauer2020, Agudo-Canalejo2019, Fruchart2021} that drive self-propelled pairs. \fixred{
In contrast, the system forms continuously ($\geq$\SI{3600}{\second}) rearranging aggregates ($N \geq 10^4$), providing a minimal and controllable platform to explore how nonreciprocity reshapes collective dynamics.
}

\begin{figure*}
    \centering
    \includegraphics{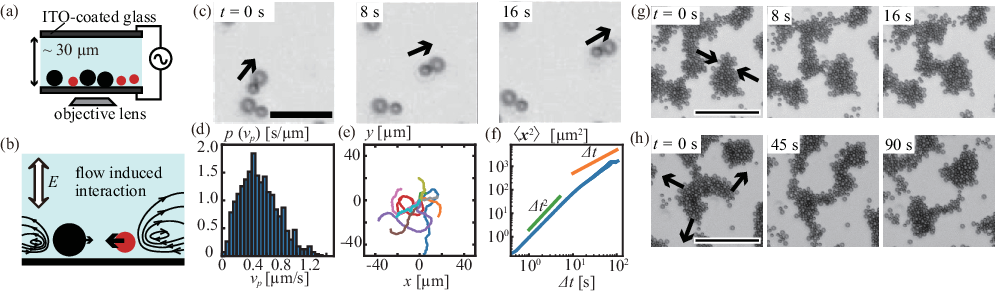}
    \caption{
    (a) Experimental setup. An in-plane AC voltage was applied; PS particles sedimented onto ITO-coated glass plates. (b) EHD flow induces size-dependent attraction, leading to nonreciprocal interactions for unequal particles. (c) Self-propelled motion of an S-L pair driven by imbalanced attraction. Scale bar: $\SI{15}{\micro\meter}$. 
    \fixred{
    (d) Velocity distribution $p(v_p)$ of 10 S-L pairs. 
    The velocity $v_p$ was obtained with time interval $\delta t$=0.5 s. 
    (e) Trajectories of the 10 S-L pairs (f) Mean square displacement $\langle \bm{x}^2 \rangle$ showing ballistic short-time behavior.
    }
    (g,h) Spontaneous accumulation and division of clusters in a bidisperse suspension (equal S and L weight density). Scale bar: $\SI{50}{\micro\meter}$.}
    \label{fig1new}
\end{figure*}

{\it Experiments-} The sample consisted of polystyrene (PS) spheres dispersed in an aqueous phase and was confined between two transparent glass electrodes coated with indium-tin oxide (ITO) [see \fixred{End Matter Sec.~\ref{EM:Exp} for details}]. This configuration allowed application of electric field parallel to the observation plane [Fig.~\ref{fig1new}(a)].


When an external field was applied, the contrast in dielectric constant between the PS particles and the surrounding water induced dipole moments in the particles. At the same time, an electric double layer (EDL), a diffuse cloud of counter-ions, formed adjacent to the electrode surfaces, with a characteristic thickness set by the Debye length.
The induced dipoles distorted the electric field near the electrodes, giving rise to tangential field components within the EDL. These tangential fields acted on the ions in the EDL and generated electroosmotic slip along the electrode surfaces, thereby driving bulk electrohydrodynamic (EHD) flow. \fixred{The EHD flow thus arises as a secondary effect of the applied electric field, and its strength scales as $|\bm{E}|^2$~\cite{Ristenpart2003, Ristenpart2004, Ristenpart2007d, Prieve2010-pv, Ma2012, Ma2015}.} Because the flow depends on the square of the field amplitude, its sign does not invert when the field polarity is reversed; in other words, the alternating field produces a rectified effect that persists under AC driving. Furthermore, the flow magnitude increases steeply with particle size, approximately as the fourth power of the particle radius [see Supplementary Information (SI)~\cite{supplementary}, Sec.~\ref{EHDcalc} for details].
\fixred{In contrast to Quincke rollers~\cite{Peters2005-mr,Bricard2013-vb}, where individual particles self-propel under DC driving, propulsion here arises only through interparticle interactions [see SI~\cite{supplementary}, Sec.~\ref{SI:Quincke}].}

As illustrated in Fig.~\ref{fig1new}(b), the strong size dependence implies that the effective EHD-mediated force between two particles can become nonreciprocal when their radii differ: the larger particle generates a faster flow, which exerts a stronger attraction on the smaller particle, thereby breaking action-reaction symmetry. 
In this study, we used bidisperse mixtures of PS spheres with radii $\SI{1}{\micro\meter}$ (``S'' particles) and $\SI{1.5}{\micro\meter}$ (``L'' particles) with equal weight density of S and L particles.
The applied AC field was $V_{pp}=\SI{5}{\volt}$ at $f=\SI{1}{\kilo\hertz}$.

\fixred{
The typical behavior of a bidisperse colloidal pair is shown in Fig.~\ref{fig1new}(c). The pair propels itself due to the imbalance of the effective mutual attraction and is transient, being disrupted by thermal fluctuations.
Such pair propulsion directly implies a violation of action–reaction symmetry, since reciprocal interactions cannot produce net center-of-mass motion of an isolated pair. The velocity distribution $p(v_p)$ [Fig.~\ref{fig1new}(d)], representative trajectories [Fig.~\ref{fig1new}(e)], and the mean square displacement $\langle \bm{x}^2\rangle$ [Fig.~\ref{fig1new}(f)] characterize this motion. All probability density functions shown are normalized to unity. Their mean speed is 0.51~\SI{}{\micro \meter \per \second}. $\langle \bm{x}^2\rangle$ shows those typical to the self-propelled particles~\cite{Peruani2007-xj} that the motion is ballistic in short timescale($t\lesssim10$~\SI{}{\second}), while it is diffusive in long timescale.
}

On larger scales, self-propelled pairs assemble into clusters [Fig.~\ref{fig1new}(g,h)], which repeatedly divide due to ongoing pair formation and separation.

\begin{figure*}
    \centering
    \includegraphics{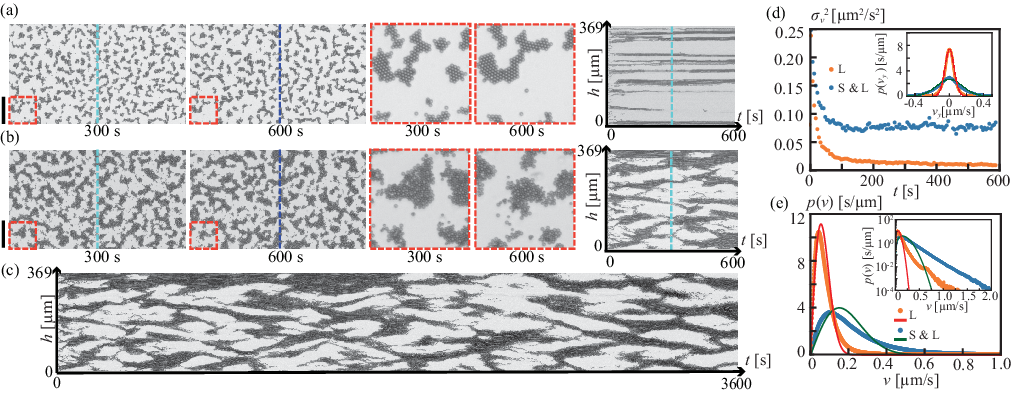}
    \caption{
    Dynamics of (a) monodisperse and (b) bidisperse suspensions. Snapshots at $t=\SI{300}{\second}$ and $\SI{600}{\second}$ are shown; spatio-temporal plots are taken along the indicated lines. Scale bar: $\SI{100}{\micro\meter}$.
    (a) L particles form static aggregates. (b) Bidisperse mixtures exhibit persistent cluster rearrangement. (c) Extended time window showing persistent dynamics up to 3600 s. 
    \fixred{
    (d) Temporal evolution of the velocity variance $\sigma_v^2$ for monodisperse(L) and bidisperse(S\&L) suspensions. Inset: velocity distributions $p(v_y)$ with Gaussian fits. (e) Speed distributions $p(v)$ with radial Gaussian fits. Fitted widths are $\sigma_v=0.054$~\SI{}{\micro\meter\per\second} (monodisperse) and $\sigma_v=0.15$~\SI{}{\micro\meter\per\second} (bidisperse).   
    }
    }
    \label{fig2}
\end{figure*}

This repeated cycle of formation and fragmentation prevents the bidisperse system from relaxing into a single static aggregate. Instead, the bidisperse aggregates continuously generate small, mobile fragments that keep the entire system in motion. Such persistent activity stands in sharp contrast to the behavior of monodisperse colloids. Figure~\ref{fig2}(a) shows snapshots and a spatio-temporal representation of aggregation in a system of L particles. In this case, the particles rapidly assemble into a crystallized, immobile cluster, and the spatio-temporal plot demonstrates that the dynamics converge almost immediately after the field is applied.

In contrast to the monodisperse case, bidisperse suspensions (S and L particles) [Fig.~\ref{fig2}(b)] form dynamically reconfiguring clusters that repeatedly fragment and merge while maintaining a finite background density. The rearrangement processes in the bidisperse system persist for up to \SI{3600}{\second} (Fig.~\ref{fig2}(c)), highlighting the striking difference in their dynamical behavior.
This qualitative difference was robust over the range of experimental parameters explored.

The impact of transient pair formation is clearly observed in Fig.~\ref{fig2}(d), which shows the time evolution of the variance of the particle-velocity field, $\sigma_v^2$ , which serves as an order parameter of collective activity. For the monodisperse suspension, $\sigma_v^2$ rapidly decays indicating the loss of active motion even though the electric field remains applied. By contrast, in the bidisperse suspension, $\sigma_v^2$ remains at finite value over the observation window, and the aggregates continue to exhibit motion through ongoing rearrangements.

\fixred{
The inset in Fig.~\ref{fig2}(d) and (e) further illustrates the difference in steady-state dynamics by showing the velocity distribution $p(v_y)$, calculated after $t=\SI{200}{\second}$ to avoid initial transients. In Fig.~\ref{fig2}(d), $p(v_y)$ is sharply peaked near zero in the monodisperse case, reflecting the immobility of the crystallized aggregates. In contrast, the bidisperse system exhibits a broader distribution with significant weight at finite velocities, confirming the sustained active motion driven by nonreciprocal pair interactions. The careful observation with $p(v)$, where $v=|\bm{v}|$ shows the existence of fat-tail behavior [See SI~\cite{supplementary} Sec.~\ref{SI:Expfit}].
}

\fixred{These observations indicate that nonreciprocal interactions prevent the usual coarsening into static aggregates and instead maintain persistent cluster dynamics.
}

\begin{figure}
    \centering
\includegraphics{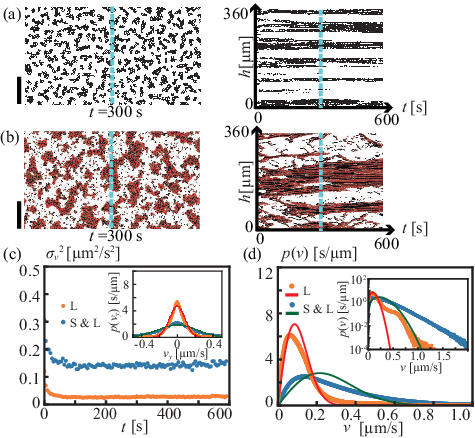}
    \caption{
    (a,b) Numerical simulations in a system of size \SI{648}{\micro\meter}$\times$\SI{360}{\micro\meter}: (a) monodisperse ($N_L=10000$) and (b) bidisperse ($N_L=5000$, $N_S=17000$). Particle radii were set to $\ell_L=s_L=\SI{1.5}{\micro\meter}$ and $\ell_S=s_S=\SI{1}{\micro\meter}$. Spatio-temporal plots are taken along the indicated line. Scale bar: $\SI{100}{\micro\meter}$.
    \fixred{(c) Temporal evolution of the velocity variance $\sigma_v^2$ for monodisperse(L) and bidisperse(S\&L) suspensions. 
    Inset: velocity distributions $p(v_y)$. 
    (d) Speed distributions $p(v)$ with Gaussian fits. 
    Fitted widths are $\sigma_v=0.086$~\SI{}{\micro\meter\per\second} (monodisperse) and 
    $\sigma_v=0.22$~\SI{}{\micro\meter\per\second} (bidisperse).}
    }
    \label{fig_sim1}
\end{figure}

{\it Agent-based model-} Various physical processes could, in principle, contribute to the observed particle dynamics. To isolate the essential mechanism underlying the behavior of bidisperse particles, we constructed a simplified mathematical model based on Brownian dynamics simulations. In this model, the system is represented by two-dimensional disks subject to thermal fluctuations, with steric exclusion and an additional effective force induced by EHD interactions~\cite{Ma2012, Ma2015}.

With appropriate nondimensionalization, we obtain the following equation,
$d\hat{\bm{x}}_i/d\hat{t}=\hat{\bm{\xi}}_i(\hat{t}) + \hat{s}_i^{-1}\sum_{j} \hat{\bm{F}}_{ij},$
where 
$\langle \hat{\xi}^\mu_i (\hat{t}) \rangle=0$,
$\langle \hat{\xi}^{\mu}_i (\hat{t}) \hat{\xi}^{\nu}_j (\hat{t'})  \rangle
= \hat{\sigma}^2 \hat{s}_i^{-1}\delta_{\mu \nu} \delta_{ij} \delta(\hat{t}-\hat{t'}).$
The total force is given by $\hat{\bm{F}}_{ij}=\hat{\bm{F}}_{ij}^{\mathrm{col}}+\hat{\bm{F}}^{\mathrm{EHD}}_{ij}$, where $\hat{\bm{F}}_{ij}^{\mathrm{col}}$ is a soft-core linear spring repulsion whose natural length is $\hat{s}_i+\hat{s}_j$, and
\begin{align}
	\hat{\bm{F}}^{\mathrm{EHD}}_{ij}=
		-\hat{\alpha} \hat{\ell}_j^4  (\hat{\bm{r}}_{ij}^2+\hat{\ell}_j^2)^{-5/2} \hat{\bm{r}}_{ij}, 
\end{align}
with a cutoff at $|\hat{\bm{r}}_{ij}|=1$, where $\hat{\bm{r}}_{ij}=\hat{\bm{x}}_{i}-\hat{\bm{x}}_{j}$.

We have separately parameterized the particle size with $\hat{s}_i$ and $\hat{\ell}_i$: the former enters the soft-core linear spring repulsion, friction, and thermal fluctuation while the latter appears in $\hat{\bm{F}}^{\mathrm{EHD}}_{ij}$. The model is nondimensionalized using the unit distance $\lambda$ and unit time $\tau$, which correspond to the cutoff distance of the EHD-induced interaction and the relaxation time of particle velocity, respectively. 
\fixred{
All parameters except the nonreciprocal interaction strength $\hat{\alpha}$ were determined from experimentally plausible values. The packing fraction of the particles was estimated from the experiment (See SI~\cite{supplementary}, Sec.~\ref{SI:Est}). The viscous damping coefficient was set as $6\pi \eta s_i$, and the noise intensity satisfies the fluctuation-dissipation theorem at 300 K. The steric repulsion strength was chosen to ensure numerical stability. The remaining parameter $\hat{\alpha}$ was adjusted to reproduce the experimentally observed propulsion speed. The qualitative phenomenology is robust against moderate variations of parameters. Detailed numerical values as well as the details of the simulation are provided in the End Matter Sec.~\ref{EM:Sim}. See also SI~\cite{supplementary}, Sec.~\ref{SI:SimPair} for the motion of a single L-S pair corresponds to Fig.~\ref{fig1new}(c,d)).
}

Results of the agent-based simulations are summarized in Fig.~\ref{fig_sim1}. The result of the numerical simulation was exemplified with dimensional quantities are shown by rescaling the nondimensional simulation variables,
so that the results can be directly compared with experiments.
The agent-based model reproduces the observed pair propulsion and persistent cluster dynamics, confirming nonreciprocal interactions as the minimal ingredient both in the single-size case [Fig.~\ref{fig_sim1}(a)], and in bidispersed case [Fig.~\ref{fig_sim1}(b)]. Semi-quantitative aggreement was also observed in the active motion in aggregate [Fig.~\ref{fig_sim1}(c)].

\fixred{Notably, in numerical simulation, $p(v)$ shows the existence of fat-tail behavior as in the experimental results (See SI~\cite{supplementary} Sec.~\ref{SI:Expfit}).
We found quantitative differences in cluster size (See SI~\cite{supplementary} Sec.~\ref{SI:ClusterSize_Area}), which arise from the absence of many-body hydrodynamic interactions in the present pairwise model.
}

The distinctive difference between the monodisperse and bidisperse systems arises from self-propulsion generated by effective nonreciprocal interactions. Furthermore, excluded-volume effects induce a divergence in the polarity of such self-propulsion, resulting in the fragmentation of clusters and the continuous active motion of aggregates. Here, we discuss these two ingredients in more detail.

Using a two-particle system with positions, the equations of motion (EOM) can be separated into those for the relative coordinate 
and those for the hydrodynamic center
(see SI~\cite{supplementary}, Sec.~\ref{SI:PairInt} for details). From the EOM for the relative coordinate, we can obtain an effective potential even when the interaction is nonreciprocal. 
A crude estimation based on a Flory-Huggins-like lattice-gas argument with the effective potential suggests that S and L particles mix after they form aggregates.

From the EOM for the hydrodynamic centers, we obtain an effective force responsible for self-propulsion, which drives the S particle toward the L particle. When the two particles are in contact, the resulting propulsion speed reaches $7.6\times10^{-4}$ in dimensionless units for $\hat{\alpha}=0.005$. This corresponds to a dimensional speed of $\SI{0.69}{\micro\meter\per\second}$, which is consistent with the experimental value (see Fig.~\ref{fig1new}(c)).

\begin{figure}
    \centering
    \includegraphics{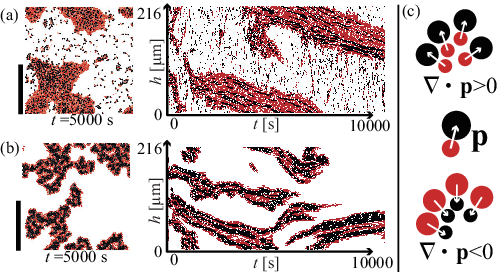}
    \caption{
    (a,b) Numerical simulations with 
    $L_x=L_y=\SI{216}{\micro\meter}$, $N_\mathrm{I}=1000$, and $N_\mathrm{II}=3000$. 
    (a) Head--large geometry ($\ell_\mathrm{I}=s_\mathrm{I}=\SI{1.5}{\micro\meter}$, $\ell_\mathrm{II}=s_\mathrm{II}=\SI{1}{\micro\meter}$). (b) Tail--large geometry (particle sizes exchanged). Scale bar: $\SI{100}{\micro\meter}$.
    (c) Schematic polarity field $\bm{p}$ illustrating $\bm{\nabla}\cdot\bm{p}>0$ (head--large) and $\bm{\nabla}\cdot\bm{p}<0$ (tail--large).
    }
    \label{fig_for_P}
\end{figure}

Thus, we find that self-propulsion induced by nonreciprocal interactions is one of the key ingredients for active clusters. However, this is not the sole factor, as revealed by agent-based numerical simulations. \fixred{To identify the geometric origin of the polarity divergence responsible for cluster fragmentation, we compare two particle arrangements: head–large and tail–large configurations.} In our experimental results (Fig.~\ref{fig1new}), S-L pairs exhibited propulsion in the head--large geometry. In the simulations, we set $\hat{\ell}_{\mathrm{I}}=\hat{s}_{\mathrm{I}}=1/6$ and $\hat{\ell}_{\mathrm{II}}=\hat{s}_{\mathrm{II}}=1/9$ to reproduce the same head--large pair structure, which resulted in continuous motion as shown in Fig.~\ref{fig_for_P}(a). Here, Roman numerals are used to denote particle types. In this configuration, the active clusters autonomously rearranged and produced small clusters. By contrast, autonomous cluster fission was suppressed when the pair geometry was inverted to a tail--large configuration. This situation was realized by setting $\hat{\ell}_{\mathrm{II}}=\hat{s}_{\mathrm{I}}=1/6$ and $\hat{\ell}_{\mathrm{I}}=\hat{s}_{\mathrm{II}}=1/9$, as shown in Fig.\ref{fig_for_P}(b) (see also SI~\cite{supplementary}, Sec.~\ref{SI:ClusterSize} for cluster-size distributions). In this geometry, the self-propelling force between large and small particles was still present, yet the formation of small clusters was strongly suppressed. These results indicate that self-propelling pairs with head-tail asymmetry, in which the head is larger than the tail, can drive autonomous cluster fission.

\fixred{
We should note that quantitative differences in cluster size arise from the absence of many-body hydrodynamic interactions in the present pairwise model (see SI~\cite{supplementary}  Sec.~\ref{SI:ClusterSize_Area}). In experiments, hydrodynamic coupling can enhance cluster growth in the monodisperse case and limit aggregate size in the bidisperse case, whereas the dry model lacks this collective regulation. 
Importantly, the qualitative emergence of persistent activity and fragmentation is reproduced without invoking hydrodynamic many-body effects.
}

\fixred{Finally, we suggest the key ingredients for the fragmentation for the cluster is that the vector field $\bm{p}$ that denotes the polarity of the self-propulsion of a pair. With the head--large geometry, positive $\nabla \cdot \bm{p}$ is favored, corresponding to the spontaneous splay-like polarity (Fig.~\ref{fig_for_P}(c)) that induces the fission of the cluster (see also SI~\cite{supplementary}, Sec.~\ref{SI:SISF}, \ref{SI:divp}).
The polarity–density coupling underlying this instability is further illustrated by a minimal continuum model (see SI~\cite{supplementary}, Sec.~\ref{SI:contmodel} for details), which shows that polarity divergence destabilizes aggregates for positive coupling.
}

{\it Summary-} In conclusion, we demonstrated that nonreciprocal interactions mediated by EHD flows profoundly influence the collective behavior of colloidal suspensions. While the monodisperse colloids form static crystalline aggregates~\cite{Ristenpart2003, Ristenpart2004, Ristenpart2007d}, bidisperse colloids generate self-propelled pairs that continuously fragment and reassemble into persistently active clusters.

Together, the agent-based and continuum models identify two key mechanisms: (i) self-propulsion arising from asymmetric EHD interactions between unequal-sized particles, and (ii) splay-like fragmentation driven by head-tail polarity due to excluded-volume effects. These mechanisms consistently account for the experimentally observed non-settling dynamics.
\fixred{
In contrast to the monodisperse particle system, where reciprocal EHD interactions lead to static aggregation, size asymmetry introduces spontaneous polarity divergence that destabilizes clusters. This nonreciprocal coupling enables persistent fragmentation and dynamic steady states despite the same underlying EHD flow.
}

\fixred{
Here, we compare our system with prefabricated dimers ~\cite{Ma2015, Niu2017, Niu2018, Wang2014-is, Shields2017-zv,Diwakar2022-qp, Diwakar2024-go}. While permanently fused dimers act as rigid self-propelled units, pairs in our system are transient and exchangeable. This exchangeability enables rapid reconfiguration and suppresses the formation of long-lived solid-like aggregates (see SI~\cite{supplementary}, Sec.~\ref{SI:dimer}). Moreover, prefabricated dimers have predominantly been investigated at small cluster sizes, 
whereas our in-situ self-assembly approach enables dense suspensions and system-spanning reconfigurable clusters ($N \geq 10^4$).
}

\fixred{
The active clusters reported in the Janus particle system~\cite{Theurkauff2012PRL} arise from the competition between motility-induced effective attraction and chemically mediated repulsion. In our system, particles interact via nonreciprocal attraction, and fragmentation is induced by excluded-volume effects associated with head-tail asymmetry.
}

Beyond clarifying the role of nonreciprocal interactions in active matter, our results suggest a simple route to engineer reconfigurable colloidal assemblies. In particular, external control of parameters such as electric field frequency~\cite{Ma2015} could enable switching between mixing
and accumulation states.
Moreover, our findings demonstrate that nonreciprocal interactions can fundamentally reprogram collective dynamics, preventing crystallization and sustaining persistent active clusters.

This work was supported by JSPS KAKENHI Grants JP16K13866, JP19H05403, JP21H01004, and JP21H00409 to YS, and JP24KJ0110 to KY as well as by JSPS and PAN under the Japan-Poland Research Cooperative Program ``Spatio-temporal patterns of elements driven by self-generated, geometrically constrained flows'', and by the JSPS Core-to-Core Program ``Advanced core-to-core network for the physics of self-organizing active matter'' (JPJSCCA20230002), to YS.

\bibliographystyle{apsrev4-2}
%

\newpage
\clearpage

\onecolumngrid
\vspace{5mm}
\begin{center}
    \large \textbf{End Matter}
\end{center}
\twocolumngrid
\section{Details of experimental method}\label{EM:Exp}
\centerline{\it Samples}

Samples consisted of aqueous suspensions of polystyrene (PS) particles. Spherical PS particles with radii $\SI{1}{\micro\meter}$ (S particles) and $\SI{1.5}{\micro\meter}$ (L particles) were used (2.5\% solids-latex; catalog nos. 19814-15 and 17134-15, PolyScience Co., Ltd.). Pure water was prepared with a Millipore Milli-Q system.

The PS particles were washed thoroughly with pure water using a microtube (CF-0150, INA-OPTIKA Inc.) to remove possible residual surfactant. As the dispersing aqueous phase, we used an aqueous solution of \SI{0.1}{\milli \mol \per \liter} NaCl and 0.2 wt.\% Pluronic F-127 (10\% in water, ANASPEC Inc.).

The overall solid fraction of PS was adjusted to approximately 1.25 wt.\% for each sample. For the monodisperse condition in Fig.~\ref{fig2}(a), suspensions containing 1.25 wt.\% of L particles only were prepared. For the bidisperse condition in Fig.~\ref{fig2}(b), suspensions were prepared by mixing S and L particles at equal weight density, i.e., 0.625 wt.\% each. Under these conditions, the packing area fraction in the field of view of Fig.\ref{fig2}(a) was approximately 30\%, corresponding to a number density of $\SI{0.0426}{\per \micro\meter\squared}$ for the L particles. In Fig.~\ref{fig2}(b), the estimated packing area fractions were 23\% and 15\% for S and L particles, respectively, corresponding to number densities of $\SI{0.10}{\per \micro\meter\squared}$ and $\SI{0.02}{\per \micro\meter\squared}$. Details of the estimation method are provided below.
\centerline{\it Chambers} 

The chambers were prepared from two glass plates coated with indium tin oxide (ITO glass, Geomatec Co., Ltd., 1007). The ITO glass was baked for 30 min at 200~\si{\degreeCelsius} in an electric furnace (MMF-1, AS ONE Inc.). The substrates were then soaked in a 10 wt.\% aqueous solution of Pluronic F-127 for 30 min, rinsed thoroughly with pure water, and dried with N$_{2}$ gas.

To assemble the chamber, two ITO glass substrates were bonded together using double-sided adhesive tape (Double-coated adhesive film No. 707, Teraoka Seisakusho Co., Ltd.). The tape thickness was \SI{30}{\micro\metre}, which defined the chamber spacing. The samples were injected into the chamber, and the openings were sealed with VALAP (a mixture of vaseline, lanolin, and paraffin in a 1:1:1 ratio) to prevent drying.

\centerline{\it Observation}

AC electric fields were applied using a function generator (WF1968, NF Corp.). Microscopic images were obtained with an IX71 inverted microscope (Evident) equipped with objective lenses (20 $\times$, LUCPLFLN20XPH, Evident). The applied field had an amplitude of $\SI{5}{\volt}$ and frequency of $\SI{1}{\kilo\hertz}$. The corresponding field intensity was approximately $1.7\times10^{5},\si{\volt\per\meter}$. Note that the electric field was always applied parallel to the observation plane.
Particle velocities $\bm{v}$ were obtained by applying dense optical flow analysis~\cite{Bradski2000} to the entire image sequence, after which velocity fields corresponding to particle-occupied regions were subsequently extracted.

\section{Detail of agent-based model and simulation method}\label{EM:Sim}
We consider an overdamped, essentially two-dimensional system of particles. Each particle $i$ has position $\bm{x}_i$ and type $\theta_i$. Two types of particles are introduced, $\theta_i=\mathrm{I}$ or $\mathrm{II}$ (where appropriate, we also use $\theta_i=L$ or $S$ to denote large and small particles, respectively). The particle radius that enters the effective EHD interaction is denoted $\ell_i$, while the radius that controls steric repulsion is denoted $s_i$. In the experiments $\ell_i$ and $s_i$ coincide, but in the simulations they were varied independently to investigate the effects of interaction imbalance and excluded-volume effects:
\begin{align}
	\ell_i=\begin{cases}
		\ell_\mathrm{I} & \theta_i = \mathrm{I}, \\
		\ell_{\mathrm{II}} & \theta_i = \mathrm{II}.
	\end{cases}
\end{align}
To avoid perfect close packing, we introduced a small size polydispersity in the steric radii $s_i$:
\begin{align}
    s_i=\begin{cases}
	s_\mathrm{I}(1+\epsilon_i) & \theta_i = \mathrm{I}, \\
	s_{\mathrm{II}}(1+\epsilon_i) & \theta_i = \mathrm{II},
    \end{cases}
\end{align}
where $\epsilon_i$ is a normally distributed random number with standard deviation $1/30$.

The equation of motion for each particle is
\begin{align}
	6\pi \eta s_i \dot{\bm{x}}_i=\bm{\xi}_i(t)+ \sum_j \bm{F}_{ij},
\end{align}
where the stochastic noise satisfies $\langle \xi^\alpha_i (t)\rangle=0$ and $\langle \xi^{\alpha}_i (t) \xi^{\beta}_j (t')  \rangle = 12\pi \eta s_i kT \,\delta_{ij} \delta_{\alpha \beta} \delta(t-t').$

The interparticle force $\bm{F}_{ij}$ includes both a soft-core linear spring repulsion $\bm{F}_{ij}^{\mathrm{col}}$ and the effective EHD attraction $\bm{F}_{ij}^{\mathrm{EHD}}$ as $\bm{F}_{ij}=\bm{F}_{ij}^{\mathrm{col}}+\bm{F}_{ij}^{\mathrm{EHD}}$ with cutoff length $\lambda$.
The soft-core linear spring repulsion is
\begin{align}
    \bm{F}_{ij}^{\mathrm{col}}=
    \begin{cases}
    k_f(s_i+s_j-|\bm{r}_{ij}|)\dfrac{\bm{r}_{ij}}{|\bm{r}_{ij}|}, & (|\bm{r}_{ij}|\leq s_i+s_j), \\
    0, & (|\bm{r}_{ij}|>s_i+s_j).
    \end{cases}
\end{align}
From Eq.~\eqref{equj2}~\cite{Ma2012, Ma2015a}, the EHD contribution is
\begin{align}\label{Eq:EHDforce}
  \bm{F}_{ij}^{\mathrm{EHD}}=
  \begin{cases}
  -\alpha \dfrac{\ell_j^4}{\big(|\bm{r}_{ij}|^2+\ell_j^2\big)^{5/2}} \bm{r}_{ij}, & (|\bm{r}_{ij}|\leq \lambda), \\
  0, & (|\bm{r}_{ij}|>\lambda).
  \end{cases}
\end{align}

We define dimensionless variables using $\lambda$ as the characteristic length and $\tau=6\pi\eta\lambda/k_f$ as the characteristic time. The dynamics then read
\begin{align}\label{eq:ndimEq}
    \dfrac{d\hat{\bm{x}}_i}{d\hat{t}}=\hat{\bm{\xi}}_i(\hat{t}) + \frac{1}{\hat{s}_i}\left[\sum_{j} \hat{\bm{F}}^{\mathrm{col}}_{ij} + \sum_{j} \hat{\bm{F}}^{\mathrm{EHD}}_{ij} \right],
\end{align}
with
\begin{align}\label{eq:ndimFcol}
    \hat{\bm{F}}_{ij}^{\mathrm{col}} &=
	\begin{cases}
		(\hat{s}_i+\hat{s}_j-|\hat{\bm{r}}_{ij}|)\dfrac{\hat{\bm{r}}_{ij}}{|\hat{\bm{r}}_{ij}|}, & (|\hat{\bm{r}}_{ij}|\leq \hat{s}_i+\hat{s}_j), \\
		0, & (|\hat{\bm{r}}_{ij}|>\hat{s}_i+\hat{s}_j).
	\end{cases} 
\end{align}
\begin{align}\label{Eq:ndimFEHD}
	\hat{\bm{F}}^{\mathrm{EHD}}_{ij} &=
	\begin{cases}
		-\hat{\alpha} \dfrac{\hat{\ell}_j^4}{(\hat{r}_{ij}^2+\hat{\ell}_j^2)^{5/2}} \hat{\bm{r}}_{ij}, & (|\hat{\bm{r}}_{ij}|\leq 1), \\
		0, & (|\hat{\bm{r}}_{ij}|>1).
	\end{cases}
\end{align}
Here $\hat{\bm{r}}_{ij}=\hat{\bm{x}}_{i}-\hat{\bm{x}}_{j}$ and $\hat{\alpha}=\frac{\alpha}{\lambda k_f}$.
The stochastic noise satisfies $\langle \hat{\xi}^\alpha_i (\hat{t}) \rangle = 0$ and $\langle \hat{\xi}^{\alpha}_i (\hat{t}) \hat{\xi}^{\beta}_j (\hat{t'})  \rangle = \frac{\hat{\sigma}^2}{\hat{s}_i} \delta_{\alpha \beta} \delta_{ij} \delta(\hat{t}-\hat{t'})$,
where $\hat{\sigma}^2 = (2kT)/(k_f \lambda^2)$, $\hat{\ell}=\ell/\lambda$, and $\hat{s}=s/\lambda$.

In typical simulations we used $\hat{\ell}_{\mathrm{I}}=\hat{s}_{\mathrm{I}}=1/6$, and $\hat{\ell}_{\mathrm{II}}=\hat{s}_{\mathrm{II}}=1/9$, corresponding to effective attractions for particles of diameter $2\ell_{\mathrm{I}}=2s_{\mathrm{I}}=\SI{3}{\micro\meter}$ and $2\ell_{\mathrm{II}}=2s_{\mathrm{II}}=\SI{2}{\micro\meter}$, with cutoff length $\lambda=\SI{9}{\micro\meter}$. In these cases, we also refer to types $\mathrm{I}$ and $\mathrm{II}$ as $L$ and $S$, respectively.  

The spring constant $k_f$ was set such that $\tau=0.01$ s, yielding $\hat{\sigma}=2.6\times10^{-3}$. The remaining parameters are the system dimensions $(\hat{L}_x,\hat{L}_y)$, the particle numbers $N_\mathrm{I}$ and $N_\mathrm{II}$, and the coupling constant $\hat{\alpha}$, which was fixed to 0.005 unless otherwise specified.

\begin{figure}
    \centering
\includegraphics{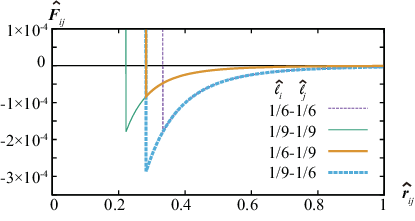}
    \caption{
    Interactions between particles $\hat{\bm{F}_{ij}}$ with different sizes. Here we set $\hat{s}_{\mathrm L}=\hat{\ell}_{\mathrm L}=1/6, \hat{s}_{\mathrm S}=\hat{\ell}_{\mathrm S}=1/9$.
    }
    \label{fig_EM_PI}
\end{figure}

\clearpage

\setcounter{equation}{0}
\setcounter{figure}{0}
\setcounter{section}{0}
\renewcommand{\thesection}{S\arabic{section}}

\renewcommand{\theequation}{S\arabic{equation}}
\renewcommand{\thefigure}{S\arabic{figure}}

\begin{center}
\textbf{\Large Supplemental Information: }
\end{center}

\section{Theoretical estimation of  EHD flows}\label{EHDcalc}

Our experiments show that two particles of different radii experience attractive interactions of different magnitudes, effectively breaking action-reaction symmetry at the level of the coarse-grained dynamics. To elucidate the mechanism underlying this non-settling behavior and to anticipate the collective dynamics in larger systems, we develop a simplified mathematical model that captures the interaction imbalance between unequal particles. The model follows the framework in Refs.~\cite{Ma2012, Ma2015}.

We first consider an effective attractive interaction generated by electro-hydrodynamic (EHD) flow. Two particles $i$ and $j$ are assumed to lie close to the bottom electrode due to gravity, with radii $\ell_i$ and $\ell_j$. Their three-dimensional positions are
$\bm{r}^{\mathrm{3D}}_i=(x_i,y_i,\ell_i)$ and $\bm{r}^{\mathrm{3D}}_j=(x_j,y_j,\ell_j)$, with in-plane projections $\bm{r}_i=(x_i,y_i)$ and $\bm{r}_j=(x_j,y_j)$.

The effective force on particle $i$ due to particle $j$ has two contributions: (i) a conservative dipole-dipole interaction that is repulsive in the lateral (2D) plane, and (ii) a nonconservative attractive component arising from EHD flow. To isolate the source of nonreciprocity, our model retains only the EHD-induced attraction and neglects the dipole-dipole term.

EHD flow is generated by the coupling between the accumulated charge in the electric double layer (EDL) at the parallel electrodes and the tangential electric field produced by the induced dipole. Under an AC field of angular frequency $\omega$, $\bm{E}_{\mathrm{ext}}(t)=E_{\mathrm{ext}}\hat{\bm{z}}\,e^{-i\omega t}$, the electrode charge within the EDL can be written as~\cite{Ristenpart2004, Ristenpart2007e, Ma2014, Ma2015, Hollingsworth2003}
\begin{align}
    Q=-\varepsilon_s \varepsilon_0 E_{\mathrm{ext}} e^{-i\omega t}\,
    \frac{\kappa H\left(1+i\omega H /\kappa D\right)}{1+\left(\omega H /\kappa D\right)^2},
\end{align}
where $\varepsilon_s$ is the solvent dielectric constant, $\varepsilon_0$ is the vacuum permittivity, $\kappa^{-1}$ is the Debye length, $2H$ is the electrode gap, and $D$ is the ionic diffusivity.

An induced dipole forms on particle $j$,
\begin{align}
	\bm{p}_j = 4\pi \varepsilon_s \varepsilon_0 \,\ell_j^3 E_{\mathrm{ext}} e^{-i\omega t} (K'+iK'')\,\hat{\bm{z}}
	\equiv p_j\,\hat{\bm{z}},
\end{align}
where $K=K'+iK''$ is the complex polarizability of PS particles~\cite{Zhou2005-yy}.

The dipole produces the electric potential
\begin{align}
\psi_j(\bm{r}^{\mathrm{3D}})=\frac{\bm{p}_j\cdot\!\left(\bm{r}^{\mathrm{3D}}-\bm{r}^{\mathrm{3D}}_j\right)}
{4\pi\varepsilon_s \varepsilon_0 \left|\bm{r}^{\mathrm{3D}}-\bm{r}^{\mathrm{3D}}_j\right|^3}
=\frac{p_j (z-\ell_j)}{4\pi\varepsilon_s \varepsilon_0 \left|\bm{r}^{\mathrm{3D}}-\bm{r}^{\mathrm{3D}}_j\right|^3}.
\end{align}
Evaluated at the electrode surface $z=0$ (i.e., $\bm{r}^{\mathrm{3D}}=(\bm{r},0)$), the tangential field is
\begin{align}
\left\{\bm{E}_j(\bm{r}^{\mathrm{3D}})\right\}_{\parallel}
&= -\left(\hat{\bm{x}}\frac{\partial \psi_j}{\partial x}+\hat{\bm{y}}\frac{\partial \psi_j}{\partial y}\right) \nonumber \\
&= -\frac{3 p_j \ell_j}{4 \pi \varepsilon_s \varepsilon_0}\,
\frac{(x\hat{\bm{x}}+y\hat{\bm{y}})}{\left|\bm{r}^{\mathrm{3D}}-\bm{r}^{\mathrm{3D}}_j\right|^5} \nonumber\\
&= -\frac{3 \ell_j^4 E_{\mathrm{ext}} e^{-i\omega t} (K'+iK'')}
{\left(|\bm{r}-\bm{r}_j|^2+\ell_j^2\right)^{5/2}}\,\bm{r}.
\end{align}

The EHD slip-driven flow due to particle $j$ scales with the product of $Q$ and the tangential field,
\begin{align}
	\bm{u}_{j}(\bm{r}) \sim \frac{\mathrm{Re}\left\{ Q \left(\bm{E}^{*}_j\right)_{\!\parallel}\right\}}
	{2 \eta \kappa}
	= \frac{\beta\, \ell_j^4}{\big(|\bm{r}-\bm{r}_j|^2+\ell_j^2\big)^{5/2}}\, \bm{r},
\label{equj}
\end{align}
where
\begin{align}
    \beta = \frac{3}{2}\,\varepsilon_s \varepsilon_0 E_{\mathrm{ext}}^2\,
    \frac{\kappa H\left\{K'+(\omega H/\kappa D)K''\right\}}{1+(\omega H/\kappa D)^2},
\end{align}
and $\eta$ is the solvent viscosity.

In the 2D geometry with $\bm{r}_{ij}=\bm{r}_i-\bm{r}_j$, we take the EHD-induced attraction on particle $i$ to be proportional to the local flow generated by particle $j$:
\begin{align}\label{equj2}
	\bm{F}^{\mathrm{EHD}}_{i\leftarrow j}(\bm{r}_i)
	\propto \frac{\ell_j^4}{\big(r_{ij}^2+\ell_j^2\big)^{5/2}}\,\bm{r}_{ij}.
\end{align}
Based on this form, we arrive at the explicit expression used in Eq.~\eqref{Eq:EHDforce}.

\fixred{Note that the flow magnitude depends on both the interparticle separation and the particle size. Because larger particles generate stronger EHD flows, taking the EHD-induced attraction to be proportional to $\bm{u}_j$ implies an imbalance of the effective attractions between unequal particles. This imbalance breaks action–reaction symmetry and generates self-propelled pairs. Although $\beta$ could in principle be estimated from experimental parameters, in Eq.~\eqref{Eq:EHDforce} we absorb this prefactor into a single coupling constant $\alpha$ and treat it as a fit parameter.
}

\fixred{
\section{Comparison with Quincke rollers}\label{SI:Quincke}
Quincke instability arises under a DC electric field for dielectric particles suspended in a weakly conducting liquid, typically low-dielectric solvents such as AOT/hexadecane mixtures. Above a critical field strength, spontaneous charge separation at the particle surface leads to a broken-symmetry state and self-propulsion via a supercritical pitchfork bifurcation~\cite{Peters2005-mr,Bricard2013-vb}. 
The instability requires a DC driving field and material parameters satisfying 
$\sigma_p/\varepsilon_p < \sigma_m/\varepsilon_m$, 
which is generally not fulfilled for particles suspended in aqueous media. 

Importantly, Quincke rollers are individually self-propelled particles, whereas in our system active motion emerges only through nonreciprocal interactions between particles of different sizes. Thus, propulsion in our experiments is intrinsically interaction-induced rather than single-particle in origin.
}

\section{Estimation of packing fraction}\label{SI:Est}

The packing fraction in the experimental data, $\varphi$ was evaluated under conditions where only L-type PS particles were dispersed in the aqueous phase. The stock suspension contained 2.5 wt.\% solid latex, which was diluted with the NaCl/Pluronic solution to the desired concentration $c$ (wt.\%). The areal fraction of particles $\bar{\varphi}$ was then observed under the AC electric field, which was necessary to attract and settle the particles onto the bottom substrate.

\begin{figure}
    \centering
    \includegraphics{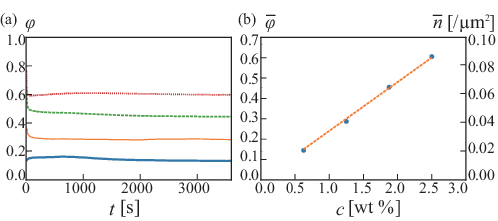}
    \caption{
    (a) Time evolution of the areal fraction $\varphi$ of PS particles with $s_L=\SI{1.5}{\micro\meter}$. The AC electric field was applied at $t=0$. The temporal average was taken over 500-3600 s to obtain $\bar{\varphi}$. The lines, from top to bottom, correspond to sample concentrations of $c=$ 2.5, 1.875, 1.25, and 0.625 wt.\%.
    (b) Dependence of the average areal fraction $\bar{\varphi}$ (left) and average number density $\bar{n}$ (right) on the sample concentration $c$. The red dashed line represents a least-squares fit.
    }
    \label{fig_PhiEst}
\end{figure}

Particle images were binarized, and $\varphi$ was calculated as
\begin{align}
\varphi = 0.907 \times \frac{S_c}{S},
\end{align}
where $S$ is the total area of the field of view and $S_c$ is the area occupied by particle clusters. The prefactor 0.907 corresponds to the packing fraction of a two-dimensional close-packed hexagonal structure. The time evolution of $\varphi$ is shown in Fig.~\ref{fig_PhiEst}(a). The temporal average over 500--3600 s yielded $\bar{\varphi}$ (Fig.~\ref{fig_PhiEst}(b)). The dependence of $\bar{\varphi}$ on concentration was fitted using a least-squares method to $\bar{\varphi}=\chi^{\varphi}_{L}\,c$, with $\chi^{\varphi}_{L}=0.240$.

The number density of L particles, $\bar{n}_L$, was calculated as
$\bar{n}_L=\bar{\varphi}/(\pi s_L^2)$, where $s_L=\SI{1.5}{\micro\meter}$ is the particle radius. 
From the fitting relation, we obtained $\bar{n}_L= \chi^{n}_{L} c$, with $\chi^{n}_{L}=0.034$ \SI{}{\per\square\micro\metre}. The number density of S particles, $\bar{n}_S$, was estimated using the scaling relation $\bar{n}_S= \chi^{n}_{S} c$, where $\chi^{n}_{S}=\chi^{n}_{L}\times(s_L/s_S)^3=0.115$ \SI{}{\per\square\micro\metre}.

Using this fitting function, the number density for the condition in Fig.~\ref{fig2}(a) ($c=1.25$ wt.\%) was estimated as $\bar{n}_L=0.0426$ \SI{}{\per\square\micro\metre}. For the condition in Fig.~\ref{fig2}(b) ($c=0.625$ wt.\% for each particle size), the number densities were $\bar{n}_L=0.0213$ and $\bar{n}_S=0.0719$ \SI{}{\per\square\micro\metre}. The field of view for these observations was $2.43\times10^5$ \SI{}{\square\micro\metre}, containing approximately 10400 L particles in Fig.~\ref{fig2}(a) and 5180 L particles and 17500 S particles in Fig.~\ref{fig2}(b).

For the numerical simulation in Fig.~\ref{fig_sim1}, the simulation domain was $72 \times 40$, corresponding to $2.32\times10^5,\si{\square\micro\meter}$. The corresponding numbers of particles were $N_L=9331$ for the monodisperse case (Fig.\ref{fig_sim1}(b)), and $(N_L, N_S)=(4665,\,15746)$ for the bidisperse case. In practice, we used $N_L=10000$ and $(N_L, N_S)=(5000,\,17000)$.

For the simulation with a domain size of $24 \times 24$ in Fig.\ref{fig_for_P}, corresponding to $4.67\times10^4$ \SI{}{\square\micro\meter}, the corresponding numbers of particles were $N_L=1868$ for the monodisperse case and $(N_L, N_S)=(934,\,3152)$ for the bidisperse case. We used $N_L=2000$ in Fig.\ref{fig_for_P}(a), and $(N_L, N_S)=(1000,\,3000)$ in Fig.~\ref{fig_for_P}(b).

\begin{figure}
    \centering
    \includegraphics{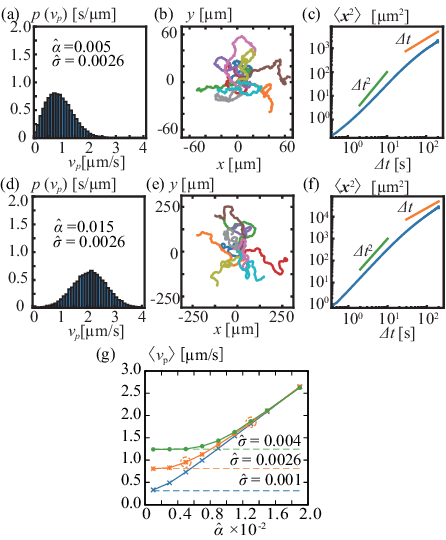}
    \caption{
    \fixred{Pair-induced motion in the agent-based simulation. Results for (a-c) $(\hat{\alpha}, \hat{\sigma})=(0.005, 0.0026)$ and (d-f) $(\hat{\alpha}, \hat{\sigma})=(0.015, 0.0026)$ are shown. 
    (a,d) Probability density function of the pair velocity, $p(v_p)$, obtained from 100 independent S-L pairs, normalized such that $\int p(v_p)\,dv_p = 1$. 
    (b,e) Trajectories of 10 representative S-L pairs. All trajectories start from the origin at $t=0$~s. 
    (c,f) Mean-squared displacement (MSD) calculated from the trajectories of all S-L pairs. 
    (g) Average velocity $\langle \hat{v}_p \rangle$ for various combinations of $(\hat{\alpha}, \hat{\sigma})$.}
    }
    \label{figSI:pairsim}
\end{figure}

\fixred{
\section{Pair-induced motion of the agent-based model}\label{SI:SimPair}

The pair-induced motion in the agent-based model was simulated based on Eq.~\eqref{eq:ndimEq}. 
We prepared $N$ independent S-L pairs. 
The particle sizes were set as $\hat{s}_S=\hat{\ell}_S=1/9$ and $\hat{s}_L=\hat{\ell}_L=1/6$, 
corresponding to $s_S=\ell_S=1~\SI{}{\micro\meter}$ and 
$s_L=\ell_L=1.5~\SI{}{\micro\meter}$ in dimensional units.

The initial positions were set to 
$\bm{x}_S=(0,0)$ and 
$\bm{x}_L=(0,1.03(s_S+s_L))$, 
so that the particles were initially separated by slightly more than contact.

The EHD interaction strength $\hat{\alpha}$ and the noise intensity $\hat{\sigma}$ were varied as control parameters. 
Each simulation was performed up to 200~s and repeated 100 times. 
The pair velocity $v_p$ was calculated from finite differences of the particle positions with a time interval $\delta t=0.5$~s.

Figures~\ref{figSI:pairsim}(a-c) show the results for 
$(\hat{\alpha}, \hat{\sigma})=(0.005, 0.0026)$, 
which are used in most simulations in the main text. 
Figures~\ref{figSI:pairsim}(d-f) show the results for 
$(\hat{\alpha}, \hat{\sigma})=(0.015, 0.0026)$, 
which are used in Figs.~\ref{figSI2} and~\ref{figSI:dimerCluster}. 
The average velocity $\langle \hat{v}_p \rangle$ for various combinations of $(\hat{\alpha}, \hat{\sigma})$ is summarized in Fig.~\ref{figSI:pairsim}(g).

The velocity distribution in Fig.~\ref{figSI:pairsim}(a) is qualitatively consistent with the experimental distribution shown in Fig.~\ref{fig1new}(d). 
However, unlike the experiment, the MSD exhibits noticeable $t$-dependence at short times for 
$(\hat{\alpha}, \hat{\sigma})=(0.005, 0.0026)$ [Fig.~\ref{figSI:pairsim}(c)], 
indicating significant positional fluctuations induced by thermal noise.

To better reproduce the experimental velocity distribution, a smaller $\hat{\alpha}$ together with a reduced $\hat{\sigma}$ would be required. 
Indeed, Fig.~\ref{figSI:pairsim}(g) shows that for $\hat{\sigma}=0.0026$, thermal fluctuations dominate the dynamics when $\hat{\alpha}\le 0.005$.

This suggests that the value $\hat{\sigma}=0.0026$, estimated from the thermal fluctuation of a spherical particle in bulk 3D aqueous media, may be overestimated for the present system. 
Since the particles are confined near a solid substrate by the electric field, the hydrodynamic drag is enhanced, which suppresses thermal fluctuations and effectively reduces the noise intensity. A quantitative estimate of the effective noise reduction near a wall would require incorporating wall-corrected hydrodynamic mobility, which is beyond the scope of the present model.
}

 \begin{figure}[tb]
    \centering  
    \includegraphics{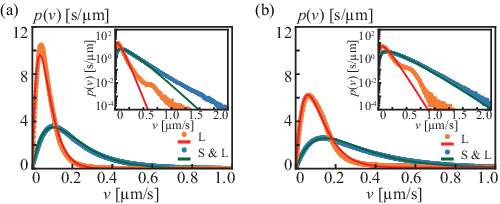}
    \caption{
    \fixred{Velocity distributions and exponential fits. (a) Experimental and (b) numerical velocity distributions $p(v)$, evaluated for $t>200$ \SI{}{\second} to exclude initial transients. Solid lines show fits to the exponential form $p(v)=(v/\Gamma^2)\exp(-v/\Gamma)$.The fitted values of $\Gamma$ are $\Gamma=0.038$~\SI{}{\micro\meter\per\second} (experiment monodisperse), $0.11$~\SI{}{\micro\meter\per\second} (experiment bidisperse), $0.060$~\SI{}{\micro\meter\per\second} (numerical monodisperse), and $0.15$~\SI{}{\micro\meter\per\second} (numerical bidisperse).
    In (b), dimensional quantities are obtained by rescaling the nondimensional simulation variables using 
    $\lambda=9$~\SI{}{\micro\meter} and 
    $\tau=0.01$~\SI{}{\second}.}
    }
    \label{exponential_fit}
\end{figure}

\fixred{
\section{Velocity distributions fitted with an exponential law}\label{SI:Expfit}

As shown in Figs.~\ref{fig2}(e) and \ref{fig_sim1}(d), the velocity distributions exhibit a pronounced long tail at large $v$. We find that both experimental and numerical distributions are better described by an exponential form than by a Gaussian. Specifically, the data are well fitted by
\begin{align}
    p(v)=\frac{v}{\Gamma^2}\exp\!\left(-\frac{v}{\Gamma}\right),
\end{align}
as shown in Fig.~\ref{exponential_fit}.

This functional form provides an excellent description of the data for both monodisperse and bidisperse systems, in experiments as well as in simulations. The observed exponential tail deviates from a Gaussian speed distribution and suggests non-Gaussian and/or intermittent driving at the level of velocity increments. At present, the microscopic origin of this behavior remains unresolved, and we therefore restrict ourselves to an empirical characterization.
}

\begin{figure}
    \centering
    \includegraphics{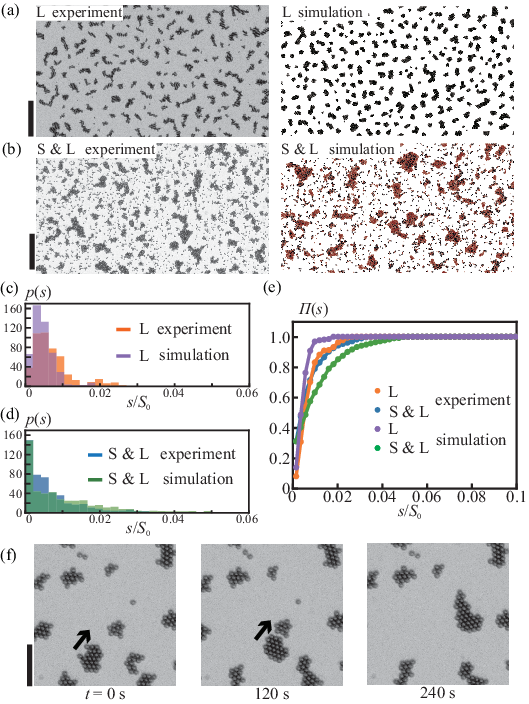}
    \caption{\fixred{Snapshots of particle configurations at $t=600$~s for (a) the monodisperse (monodisperse) system and (b) the bidisperse system. (c,d) Weighted cluster-size distribution $p(s)$ for the monodisperse and bidisperse systems, respectively, where $s$ denotes the area of an observed cluster and $S_0$ is the total particle-occupied area within the observation region. (e) Cumulative distribution function $\Pi(s)$ corresponding to (c,d). (f) Example of persistent cluster motion observed experimentally. Scale bars: (a,b) 100~\SI{}{\micro\meter}; (f) 30~\SI{}{\micro\meter}.}
    }
    \label{fig_clsize}
\end{figure}

\fixred{
\section{Cluster size analysis--Area-based measurement}\label{SI:ClusterSize_Area}
We quantified cluster sizes in experiments and numerical simulations using the occupied area. 
To avoid system-spanning percolation, the particle number was reduced to half of that used in Figs.~\ref{fig2} and~\ref{fig_sim1}. 
Snapshots were analyzed over the time window $t=200$-600~\SI{}{\second}, 
with a frame rate of 2~\SI{}{\hertz}, in order to exclude transient effects from the initial configuration.

Let $S_0$ denote the total particle-occupied area, averaged over the same time interval. 
From image analysis, we obtained the cluster-size distribution $p_s(s)$ based on the occupied area $s$. 
We then defined
\begin{align}
    p(s)=\frac{s}{S_0}p_s(s),
\end{align}
which is proportional to the probability of finding a particle within a cluster of size $s$ for a monodisperse system. For the bidisperse system, $p(s)$ does not strictly coincide with the particle-based probability, because the area contribution per particle differs between S and L particles.

Figures~\ref{fig_clsize}(a) and (b) show $p(s)$ for the monodisperse (L) and bidisperse (S\&L) systems, respectively, and the corresponding cumulative distributions are summarized in Fig.~\ref{fig_clsize}(c).

In the monodisperse system, numerical simulations produce smaller clusters compared with the experiment. Experimentally, clusters are often anisotropic and exhibit persistent translational motion, whereas such persistent motion is absent in the simulations. The persistent motion enhances inter-cluster collisions and promotes further aggregation, leading to larger clusters in the experiment. This behavior is likely associated with many-body hydrodynamic interactions, which are neglected in the present agent-based model where interactions are restricted to pairwise forces.

In contrast, in the bidisperse system, simulations yield larger clusters than those observed experimentally. Experimentally, clusters are typically anisotropic and elongated, while simulated clusters remain comparatively symmetric and continue to grow. In the simulations, particles inside a cluster continue to attract surrounding particles through pairwise interactions. However, in a real fluid system with volume conservation, hydrodynamic flow fields should saturate beyond a certain cluster size, thereby limiting further growth. The absence of this saturation mechanism in the present model likely explains the overestimation of cluster size.

This limitation reflects the absence of long-range hydrodynamic screening and mobility reduction inside dense clusters in the current pairwise model. Incorporating many-body hydrodynamic interactions into the numerical framework would therefore be an important direction for bridging the gap between experiment and simulation. 

} 

\section{Pair interaction of two particles}\label{SI:PairInt}
We next analyze the dynamics of a pair of particles. In particular, we show that the motion can be separated into an effective pseudo-potential force and an additional self-propelling component when the two particles have different radii. Throughout this section all variables are dimensionless. The soft-core linear spring repulsion is neglected, as collisions are assumed to be avoided by hard-core exclusion. We also set $\hat{s}=\hat{\ell}$ for simplicity.

Consider two particles at positions $\hat{\bm{x}}_1$ and $\hat{\bm{x}}_2$, with equations of motion
\begin{align}
	\dot{\hat{\bm{x}}}_{1} &= \hat{\bm{\xi}}_{1}(t) -\hat{\alpha} \frac{\hat{\ell}_{2}^{4}}{\hat{\ell}_1} \frac{\hat{\bm{r}}}{(\hat{r}^{2}+\hat{\ell}_{2}^{2})^{\frac{5}{2}}}, \\
	\dot{\hat{\bm{x}}}_{2} &= \hat{\bm{\xi}}_{2}(t) +\hat{\alpha} \frac{\hat{\ell}_{1}^{4}}{\hat{\ell}_2} \frac{\hat{\bm{r}}}{(\hat{r}^{2}+\hat{\ell}_{1}^{2})^{\frac{5}{2}}},
\end{align}
where $\hat{\bm{r}}=\hat{\bm{x}}_{1}-\hat{\bm{x}}_{2}$.
We define a hydrodynamic center $\hat{\bm{R}}=(\hat{\ell}_1\hat{\bm{x}}_{1}+\hat{\ell}_2\hat{\bm{x}}_{2})/(\hat{\ell}_1+\hat{\ell}_2)$.

The relative motion can then be written as
\begin{align}
	\frac{d\hat{\bm{r}}}{d\hat{t}}=\hat{\bm{\Gamma}}_{\hat{r}}+\frac{\hat{\bm{f}}}{\hat{\mathcal{L}}},
\end{align}
with
\begin{align}
	\hat{\bm{f}}=-\frac{\hat{\alpha}\hat{\bm{r}}}{\hat{\ell}_1+\hat{\ell}_2} \left[\frac{\hat{\ell}_1^5}{(\hat{r}^2+\hat{\ell}_1^2)^{\frac{5}{2}}}+\frac{\hat{\ell}_2^5}{(\hat{r}^2+\hat{\ell}_2^2)^{\frac{5}{2}}}\right],
\end{align}
and noise correlations $\langle \hat{\Gamma}^{\alpha}_{\hat{r}}(\hat{t})\rangle=0$, $\langle \hat{\Gamma}^{\alpha}_{\hat{r}}(\hat{t}) \hat{\Gamma}^{\beta}_{\hat{r}}(\hat{t}') \rangle=\frac{\hat{\sigma}^2}{\hat{\mathcal{L}}} \delta_{\alpha \beta} \delta(\hat{t}-\hat{t}')$,
where $\hat{\mathcal{L}}= (\hat{\ell}_1 \hat{\ell}_2)/(\hat{\ell}_1+\hat{\ell}_2)$.

From these expressions, we define the effective potential $\Delta \hat{u}(\hat{r}, \hat{\ell}_1, \hat{\ell}_2)= -\int_{+\infty}^{\hat{r}} \hat{\bm{f}}(\hat{\bm{r}}', \hat{\ell}_1, \hat{\ell}_2)\cdot d\hat{\bm{r}}'$.

Evaluating this integral yields
\begin{align}
	\Delta \hat{u}(\hat{r}, \hat{\ell}_1, \hat{\ell}_2)=-\frac{\hat{\alpha}}{3(\hat{\ell}_1+\hat{\ell}_2)} \left[\frac{\hat{\ell}_1^5}{(\hat{r}^2+\hat{\ell}_1^2)^{\frac{3}{2}}}+\frac{\hat{\ell}_2^5}{(\hat{r}^2+\hat{\ell}_2^2)^{\frac{3}{2}}}\right].
\end{align}
For a bound pair at contact, $\hat{r}=\hat{\ell}_1+\hat{\ell}_2$, we define $\hat{u}_{\ell_1, \ell_2} =\Delta \hat{u}(\hat{\ell}_1+\hat{\ell}_2, \hat{\ell}_1, \hat{\ell}_2)$.
For $\hat{\alpha}=0.005$, the potential depths are
$\Delta \hat{u}_{1/6,\,1/6}= -2.5 \times 10^{-5}$, 
$\Delta \hat{u}_{1/6,\,1/9}= -2.6\times 10^{-5}$, and 
$\Delta \hat{u}_{1/9,\,1/9}= -1.7 \times 10^{-5}$.  
$\Delta \hat{u}_{1/6,\,1/6}+\Delta \hat{u}_{1/9,\,1/9}-2\Delta \hat{u}_{1/6,\,1/9}=1.2 \times 10^{-5}>0$, 
indicating that particles of different sizes tend to mix within aggregates.

For the hydrodynamic center we have
\begin{align}
	\frac{d\hat{\bm{R}}}{d\hat{t}}=\hat{\bm{\Gamma}}_{\hat{R}}+\frac{\hat{\bm{F}}}{\hat{\ell}_1+\hat{\ell}_2},
\end{align}
with
\begin{align}
	\hat{\bm{F}}=\hat{\alpha}\hat{\bm{r}} \left[\frac{\hat{\ell}_1^4}{(\hat{r}^2+\hat{\ell}_1^2)^{\frac{5}{2}}}-\frac{\hat{\ell}_2^4}{(\hat{r}^2+\hat{\ell}_2^2)^{\frac{5}{2}}}\right],
\end{align}
and $\langle \hat{\Gamma}^{\alpha}_{\hat{R}}(\hat{t})\rangle=0$, $\langle \hat{\Gamma}^{\alpha}_{\hat{R}}(\hat{t}) \hat{\Gamma}^{\beta}_{\hat{R}}(\hat{t}') \rangle=\frac{\hat{\sigma}^2}{\hat{\ell}_1+\hat{\ell}_2} \delta_{\alpha \beta} \delta(\hat{t}-\hat{t}')$.

Writing $\hat{\bm{r}}=\hat{r}\bm{e}_{\theta}=\hat{r}(\bm{e}_x \cos\theta +\bm{e}_y \sin\theta)$, the relative dynamics are
\begin{align}
	\frac{d\hat{r}}{d\hat{t}}=\hat{\Gamma}^{\hat{r}}_{\hat{r}}+\frac{\hat{f}}{\hat{\mathcal{L}}}, \qquad
	\hat{r}\frac{d\theta}{d\hat{t}}=\hat{\Gamma}^{\theta}_{\hat{r}}.
\end{align}
For a bound pair, $\hat{r}=\hat{\ell}_1+\hat{\ell}_2$, we have
\begin{align}
	\frac{d\theta}{d\hat{t}}=\frac{1}{\hat{\ell}_1+\hat{\ell}_2}\hat{\Gamma}^{\theta}_{\hat{r}}\equiv \hat{\Gamma}_{\theta},
\end{align}
with correlation $\langle \hat{\Gamma}_{\theta}(\hat{t})\rangle=0$, $\langle \hat{\Gamma}_{\theta}(\hat{t}) \hat{\Gamma}_{\theta}(\hat{t}') \rangle=\frac{\hat{\sigma}^2}{\hat{\ell}_1\hat{\ell}_2(\hat{\ell}_1+\hat{\ell}_2)}  \delta(\hat{t}-\hat{t}')$.

Thus, the orientation $\bm{e}_{\theta}=(\bm{e}_x \cos\theta +\bm{e}_y \sin \theta )$ fluctuates in time, and the center position evolves as
\begin{align}
	\frac{d\hat{\bm{R}}}{d\hat{t}}=\hat{\bm{\Gamma}}_{\hat{R}}+\hat{F}_{\mathrm{VM}}\bm{e}_{\theta},
\end{align}
with effective propulsion
\begin{align}
	\hat{F}_{\mathrm{VM}}=  
    \hat{\alpha}\left[\frac{\hat{\ell}_1^4}{((\hat{\ell}_1+\hat{\ell}_2)^2+\hat{\ell}_1^2)^{\frac{5}{2}}}-\frac{\hat{\ell}_2^4}{((\hat{\ell}_1+\hat{\ell}_2)^2+\hat{\ell}_2^2)^{\frac{5}{2}}}\right].
\end{align}
This is essentially equivalent to the Vicsek model with a fixed speed. For $\hat{\alpha}=0.005$ and radii $1/6$ and $1/9$, the speed is $7.6\times10^{-4}$. 

Increasing the external electric field by a factor of $1.7$ would raise the effective coupling to $\hat{\alpha}=0.015$. In this case, the propulsion speed becomes $2.3\times10^{-3}$.

\section{Cluster size analysis--particle based measurement}\label{SI:ClusterSize}
\begin{figure}
    \centering
    \includegraphics{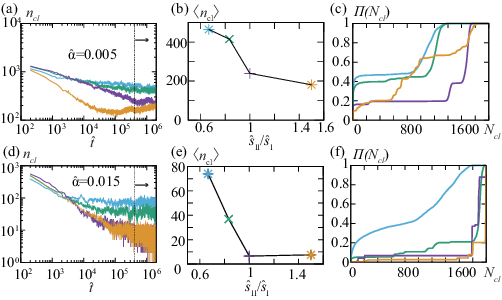} 
    \caption{
    \fixred{
    Cluster-size analysis of the agent-based model defined by Eq.~\eqref{eq:ndimEq} in a domain $\hat{L}_x=\hat{L}_y=24$ with $N_\mathrm{I}=N_\mathrm{II}=1000$. Results are shown for $\hat{\alpha}=0.005$ (a-c) and $\hat{\alpha}=0.015$ (d-f). The EHD interaction parameter were fixed at $(\hat{\ell}_\mathrm{I},\hat{\ell}_\mathrm{II})=(1/6,\,1/9)$.
    (a,d) Time evolution of the number of clusters $n_{cl}$. The steric radii were varied as $(\hat{s}_\mathrm{I},\hat{s}_\mathrm{II})=(1/6,\,1/9)$ (cyan), $(1/6,\,1/7.2)$ (green), $(1/6,\,1/6)$ (purple), and $(1/9,\,1/6)$ (orange). 
    Statistical analysis was performed using data for $\hat{t} > 4\times10^5$ (dot-dashed line).
    (b,e) Time-averaged number of clusters $\langle n_{cl} \rangle$. For $\hat{\alpha}=0.015$, complete aggregation is observed for $\hat{s}_\mathrm{II}/\hat{s}_\mathrm{I}=1$ and $1.5$.
    (c,f) Cumulative cluster-size distribution $\Pi(N_{cl})=\sum_{k=1}^{N_{cl}} p(k)$, where $p(N_{cl})$ is the probability that a particle belongs to a cluster of size $N_{cl}$. Color coding is identical to (a,d).
    }
    }
    \label{figSI2}
\end{figure}

\fixred{
To quantify the role of excluded-volume contrast, we statistically analyzed the number and size of clusters. Simulations were performed in a domain $\hat{L}_x=\hat{L}_y=24$ with $N_\mathrm{I}=N_\mathrm{II}=1000$. The EHD interaction radii were fixed at $(\hat{\ell}_\mathrm{I},\hat{\ell}_\mathrm{II})=(1/6,\,1/9)$, while the steric radii were systematically varied. This protocol maintains a constant nonreciprocal attraction while tuning the excluded-volume asymmetry.

Figure~\ref{figSI2}(a,d) shows the temporal evolution of the cluster number $n_{cl}$, which decreases and eventually reaches a steady state. The time-averaged value $\langle n_{cl} \rangle$ is summarized in Fig.~\ref{figSI2}(b,e).

For the head--large case [$(\hat{s}_\mathrm{I},\hat{s}_{\mathrm{II}})=(1/6,\,1/9)$; cyan], $\langle n_{cl} \rangle$ remains of order $10^{2}$, indicating fragmentation into many small clusters. In contrast, in the equal-size [$(1/6,\,1/6)$; purple] and tail--large [$(1/9,\,1/6)$; orange] cases, $\langle n_{cl} \rangle$ is reduced to order unity, signifying near-complete aggregation.

The cluster-size statistics were further quantified via $p(N_{cl})$, the probability that a particle belongs to a cluster 
of size $N_{cl}$. The cumulative distribution $\Pi(N_{cl})$ is shown in Fig.~\ref{figSI2}(c,f). For the equal-size and tail--large cases, $\Pi(N_{cl})$ rapidly approaches unity only near $N_{cl}\approx2000$, indicating that most particles reside in a single macroscopic cluster. By contrast, in the head--large and intermediate cases, $\Pi(N_{cl})$ increases gradually, reflecting a broad distribution of finite clusters.

These results demonstrate that excluded-volume asymmetry controls the stability of macroscopic aggregation. In particular, enhanced polarity divergence in self-propelled pairs promotes cluster fission and prevents system-spanning collapse.
}

\begin{figure}[tb]
    \centering
    \includegraphics{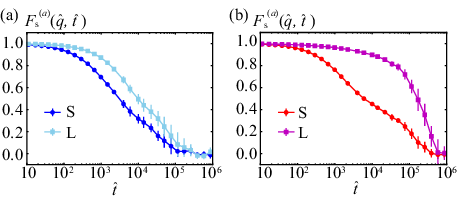}
    \caption{
    \fixred{Self part of the intermediate scattering function $F^{(a)}_s(\hat{q},\hat{t})$ for (a) $(\hat{s}_\mathrm{I},\hat{s}_\mathrm{II})=(1/6,1/9)$ and (b) $(\hat{s}_\mathrm{I},\hat{s}_\mathrm{II})=(1/9,1/6)$. Error bars represent the standard deviation over 10 time windows. S and L denote particles with $\hat{s}=1/9$ and $\hat{s}=1/6$, respectively.}
    }
    \label{fig_SN_FS}
\end{figure}

\fixred{
\section{Structural relaxation probed by the intermediate scattering function}\label{SI:SISF}

To characterize the persistent dynamics more rigorously, we examine the structural relaxation using the self-intermediate scattering function based on the data used for Fig.~\ref{figSI2}(a-c). The data was obtained after the annealing upto $\hat{t}=2\times10^6$.

The self-intermediate scattering function for particle type $a$ is defined as
\begin{align}
F_s^{(a)}(\hat{q},\hat{t})
=
\left\langle 
\frac{1}{N_a}
\sum_{i=1}^{N_a}
\exp
\left[
i\bm{\hat{q}}\cdot
(\bm{\hat{x}}_i(t)-\bm{\hat{x}}_i(0))
\right]
\right\rangle ,
\end{align}
where $\bm{\hat{q}}$ is the wave vector, $\hat{q}=|\bm{\hat{q}}|$ its magnitude, $a=\mathrm{I,II}$ denotes the particle type, and $N_a$ denotes the number of particles of that species. 
Angular brackets represent an ensemble average over time origins. To probe structural relaxation at the particle scale, we choose the wave number as
\begin{align}
\hat{q} = \frac{2\pi}{\hat{s}_a},
\end{align}
where $\hat{s}_a$ is the characteristic particle size.

Figure~\ref{fig_SN_FS}(a) shows $F_s^{(a)}(\hat{q},\hat{t})$ for the two geometries 
$(\hat{s}_\mathrm{I},\hat{s}_\mathrm{II})=(1/6,1/9)$ and $(1/9,1/6)$.
In the head--large case $[(\hat{s}_\mathrm{I},\hat{s}_\mathrm{II})=(1/6,1/9)]$, 
the relaxation time (defined as the time at which $F_s^{(a)}(\hat{q},\hat{t})$ decays to $1/e$) 
is of order $10^2$ for both particle types.
By contrast, in the tail--large case $[(\hat{s}_\mathrm{I},\hat{s}_\mathrm{II})=(1/9,1/6)]$, 
the relaxation is significantly slower than in the head--large case, and the relaxation time differs between the two particle types. 
Notably, simply exchanging the head and tail particle sizes changes the relaxation time by approximately one order of magnitude.

Combined with the cluster-size analysis in Sec.~\ref{SI:ClusterSize}, 
these results indicate that a larger polarity divergence leads to more persistent dynamics and enhanced cluster destabilization, including fission. We also note a weak bump in $F_s^{(a)}(\hat{q},\hat{t})$, which may be associated with transient cluster rotation and contact-induced reversible displacements. This analysis provides a quantitative measure of the persistence of the cluster dynamics.
}

\begin{figure}[tb]
    \centering
    \includegraphics{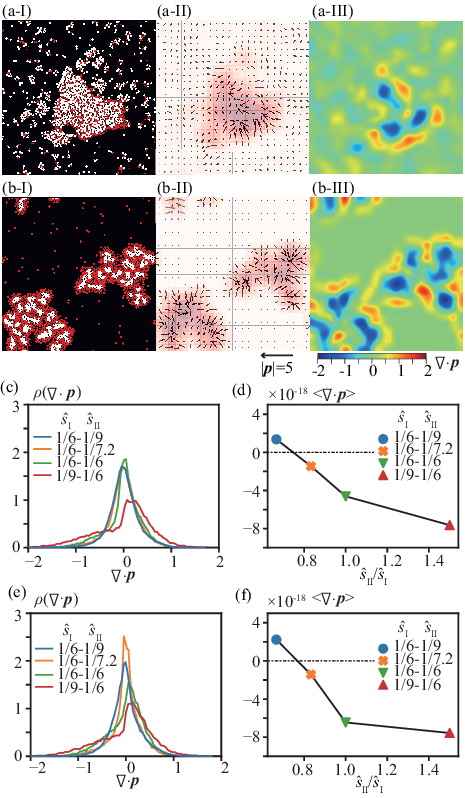}
    \caption{\fixred{(a), (b) (I) Original image of particle position ($\hat{t}=1.8\times 10^6$) and obtained (II) polarity field $\bm{p}$ and (III) its divergence $\mathrm{div} \bm{p}$ for $\hat{\alpha}=0.005$ and $\hat{s}_\mathrm{II}/\hat{s}_\mathrm{I}=$(a)0.667 and (b) 1.5. The length of arrow for (II) and colormap for (III) are indicated at the bottom of (b). (c,e) Probability distribution $\rho(\mathrm{div})$ for $\hat{\alpha}=$ (c) 0.005 and (e) 0.015 with different combination of  $\hat{s}_\mathrm{I}$ and $\hat{s}_\mathrm{II}$ are shown. Probability distribution is normalized to be unity when it is integrated over the possible $\mathrm{div}\bm{p}$. (d,f) The mean valued of $\mathrm{div}\bm{p}$ plotted with $\hat{s}_\mathrm{II}/\hat{s}_\mathrm{I}$ for  $\hat{\alpha}=$ (d) 0.005 and (f) 0.015.}
    }
    \label{figSI:divp}
\end{figure}

\fixred{
\section{Measurement of divergence of polarity}\label{SI:divp}
To quantify the spatial organization of self-propelled pairs, we measured the divergence of the coarse-grained polarity field, as shown in Fig.~\ref{figSI:divp}. 

Simulation data were based on the data used for Fig.~\ref{figSI2}(a-c). The data were sampled every $\delta\hat{t}=200$ up to $\hat{t}=2\times10^6$. The EHD interaction parameter were fixed at $(\hat{\ell}_\mathrm{I},\hat{\ell}_\mathrm{II})=(1/6,\,1/9)$.

For each bonded I-II pair, we defined a unit vector $\bm{n}_i$ at position $\bm{x}_i$ pointing from particle II to I. This direction corresponds to the net (unbalanced) force that drives the propulsion of the pair. A pair was identified when the center-to-center distance satisfied $|\hat{\bm{r}}_{ij}|\le 1.5(\hat{s}_\mathrm{I}+\hat{s}_\mathrm{II})$.

The coarse-grained polarity field $\bm{p}_{k,l}$ was constructed on a square mesh $\Delta_{k,l}$ of size $1\times1$ as
\begin{align}
\bm{p}_{k,l}=\sum_{i,\,\bm{x}_i\in\Delta_{k,l}} \bm{n}_i.
\end{align}
To reduce fluctuations, we performe 
a spatial smoothing over the nearest and next-nearest neighboring cells.
Typical polarity field are shown in Fig.~\ref{figSI:divp}(a-II) and (b-II), with the corresponding particle configurations in (a-I) and (b-I). In both geometries, the polarity field exhibits convergent structures. However, for $\hat{s}_\mathrm{II}/\hat{s}_\mathrm{I}=1.5$ (tail--large geometry), the convergent pattern is significantly enhanced, corresponding to the formation of dense clusters.

The discrete divergence was evaluated as
\begin{align}
(\nabla\cdot\bm{p})_{k,l}
=
\frac{p_{x,k+1,l}-p_{x,k-1,l}}{2\Delta}
+
\frac{p_{y,k,l+1}-p_{y,k,l-1}}{2\Delta},
\end{align}
with $\Delta=1$.

Typical divergence fields are shown in Fig.~\ref{figSI:divp}(a-III) and (b-III). Negative divergence is observed in the interior of clusters, while positive divergence appears near the periphery.

The probability distribution $\rho(\nabla\cdot\bm{p})$ is shown in Fig.~\ref{figSI:divp}(c,e) for $\hat{\alpha}=0.005$ and $0.015$. A broader distribution is observed in the tail--large geometry, consistent with stronger polarity accumulation in dense clusters.

The mean value $\langle \nabla\cdot\bm{p} \rangle$ is plotted as a function of $\hat{s}_\mathrm{II}/\hat{s}_\mathrm{I}$ in Fig.~\ref{figSI:divp}(d,f). For both values of $\hat{\alpha}$, the mean divergence decreases monotonically as $\hat{s}_\mathrm{II}/\hat{s}_\mathrm{I}$ increases. In particular, $\langle \nabla\cdot\bm{p} \rangle$ is positive 
for$\hat{s}_\mathrm{II}/\hat{s}_\mathrm{I}=0.667$ and approaches large negative values in the tail--large regime.

These results indicate that steric asymmetry systematically controls the sign and magnitude of the polarity divergence. Positive divergence corresponds to outward-directed polarity, which enhances effective pair propulsion, whereas negative divergence reflects polarity accumulation within dense clusters.
}

\section{Continuum hydrodynamic description}\label{SI:contmodel}

\fixred{
Both experiments and agent-based simulations indicate that collective motion of clusters originates from (i) imbalance of effective attraction between particles of different sizes and (ii) divergent polarity generated by asymmetric pairs (head larger than tail). To test this mechanism at a coarse-grained level, we construct a minimal continuum model based on a conserved density field $\rho$ and a non-conserved polarity field $\bm{p}$.

A central ingredient of the continuum description is the emergence of polar order associated with self-propelled pairs. 
Propulsion arises only when a finite local density of pairs is present. 
We therefore describe the system in terms of a conserved scalar density field $\rho(\bm{r},t)$ representing the total colloid density, 
and a non-conserved polar field $\bm{p}(\bm{r},t)$ representing the coarse-grained polarity of pairs.

Since $\rho$ is conserved whereas $\bm{p}$ is not, the dynamics are written as
\begin{align}
\partial_t \rho + v_0 \nabla\cdot(\rho\bm{p})
= \frac{1}{\gamma_\rho} \nabla^2 \frac{\delta \mathcal{F}}{\delta \rho},
\end{align}
\begin{align}
\partial_t \bm{p}
+\lambda_1 (\bm{p}\cdot\nabla)\bm{p}
= -\frac{1}{\gamma} \frac{\delta \mathcal{F}}{\delta \bm{p}},
\end{align}
where $\mathcal{F}$ is an effective free-energy functional.

We model the mixture of S and L particles by a single density field $\rho$, consistent with the effective mixing discussed in Sec.~\ref{SI:PairInt}. This approximation is justified because the effective interaction leads to mixed aggregates of S and L particles, and no macroscopic demixing between species is observed in either experiments or agent-based simulations. Density segregation driven by effective EHD attraction is incorporated through a Cahn--Hilliard-type free energy,
\begin{align}
\mathcal{F} = & 
\int d^2\bm{r}
\left[ 
\frac{\alpha}{4}\rho^2(\rho-\rho_0)^2
+\frac{\kappa}{2} (\nabla \rho)^2 \right. \nonumber \\
& \left. +\frac{1}{2}\zeta |\mathbf{p}|^2
-w\, \rho\, (\nabla\cdot\bm{p})
\right].
\label{eq_main_rho}
\end{align}
The first two terms describe phase separation into 
low- and high-density regions, 
with coexistence between $\rho=0$ and $\rho=\rho_0$ when $w=0$, 
corresponding to the monodisperse system. 
The third term is a local polarity relaxation term, which ensures the finite lifetime of pair-induced polarity in the experiment.
The fourth term couples the density field to the divergence of the polarity.
For $w>0$, configurations with positive $\nabla\cdot\bm{p}$ are energetically favored, 
leading to spontaneous splay-like polarity consistent with the head--large geometry observed in particle-based simulations. This minimal coupling isolates the role of polarity divergence as the mechanism that destabilizes dense aggregates.
}

\begin{figure}[tb]
    \centering
    \includegraphics{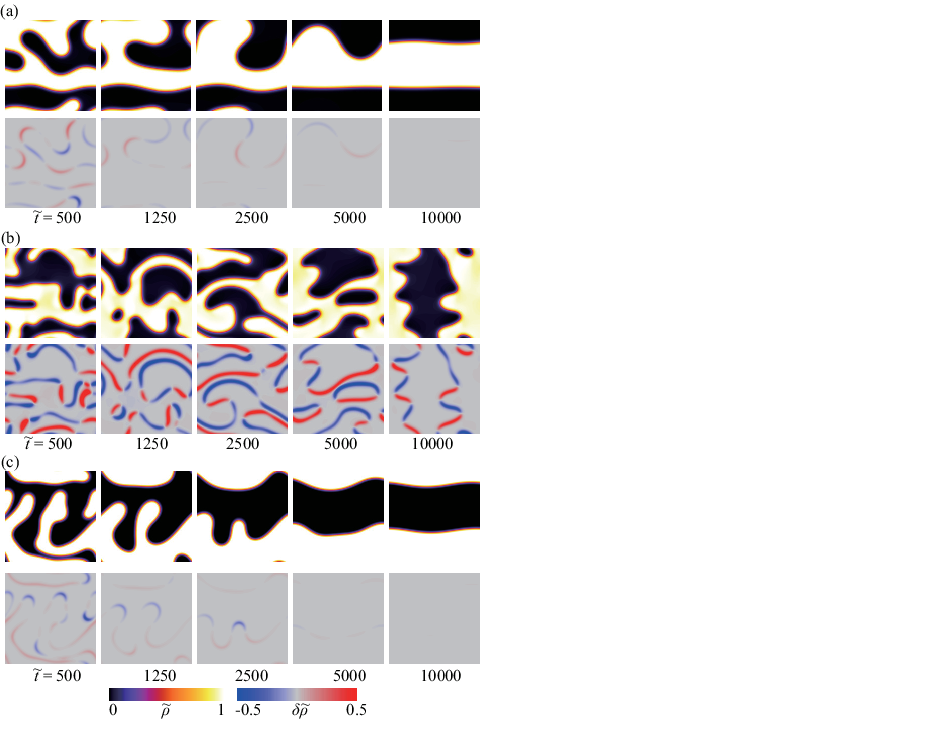}
    \caption{\fixred{Behavior of the continuum model based on Eq.~\eqref{eq_main_rho} for $\tilde{w}$ is (a) 0 (b) 0.001 and (c)-0.001. The top panels show snapshots of the density field $\tilde{\rho}$, while the bottom panels display the temporal difference $\delta \tilde{\rho}$ computed between configurations separated by a time interval $\Delta \tilde{t}=25$. 
    }
    }
    \label{figSI:cont}
\end{figure}

\fixred{
From the free-energy functional, we obtain
\begin{align}
\partial_t \rho  &+v_{0} \nabla \cdot(\rho \bm{p})  \nonumber \\
=& \frac{1}{\gamma_{\rho}} \nabla^{2}\left[\alpha \rho\left(\rho-\rho_{0}\right)\left(\rho-\frac{\rho_{0}}{2}\right)-\kappa \nabla^{2} \rho-w(\nabla \cdot \bm{p})\right],
\end{align}

and
\begin{align}
\partial_t \bm{p}
=
-\frac{\zeta}{\gamma}\bm{p}-\frac{w}{\gamma}\,\nabla \rho .
\end{align}
We neglect the nonlinear convective term $\lambda_{1}(\bm{p}\cdot\nabla)\bm{p}$, as it is higher order in $\bm{p}$ and does not affect the leading-order mechanism discussed here. Note that the dynamics are not purely relaxational because the density equation contains the advective flux $v_0\nabla\cdot(\rho\bm{p})$, which breaks detailed balance even if $\mathcal F$ is introduced as a convenient coarse-grained functional.

We also emphasize that the present formulation is intended to capture the leading-order effect of the polarity divergence coupling. Accordingly, we restrict our attention to sufficiently small values of $w$, so that this coupling acts as a perturbative correction to the Cahn--Hilliard dynamics. In this regime, polarity is generated primarily at density gradients, i.e., at aggregate interfaces.

Here we introduce dimensionless variables by using $\rho_0$ as the density unit, 
$\alpha \rho_0^2/(v_0\gamma_\rho)$ as the length unit, and 
$\alpha \rho_0^2/(v_0^2\gamma_\rho)$ as the time unit. 
Here $\bm{p}$ is dimensionless. 
The resulting equations are
\begin{align}
\frac{\partial \tilde{\rho}}{\partial \tilde{t}}
+&\tilde{\nabla}\cdot(\tilde{\rho}\,\bm{p})
\nonumber \\ =
&\tilde{\nabla}^{2}\left[\tilde{\rho}(\tilde{\rho}-1)\left(\tilde{\rho}-\tfrac{1}{2}\right)
-\tilde{\kappa}\tilde{\nabla}^{2} \tilde{\rho}
-\tilde{w}\, \tilde{\nabla} \cdot \bm{p}\right],\label{eq:contndim1}
\end{align}
and
\begin{align}
\frac{\partial \bm{p}}{\partial \tilde{t}}
&=-\frac{\tilde{\zeta}}{\tilde{\gamma}}\bm{p}-\frac{\tilde{w}}{\tilde{\gamma}} \tilde{\nabla} \tilde{\rho},\label{eq:contndim2}
\end{align}
where tildes denote dimensionless quantities and
\begin{align}
\tilde{\kappa}=\kappa \frac{v_0^2 \gamma_\rho^2}{\alpha^3 \rho_0^6},\, 
\tilde{\zeta}=\zeta\frac{1}{\alpha\rho_0^3},\,
\tilde{w}=w \frac{v_0\gamma_\rho}{\alpha^2\rho_0^5}, \,
\tilde{\gamma}=\gamma \frac{v_0^2\gamma_\rho}{\alpha^2 \rho_0^5}.
\end{align}

For the numerical simulations shown in Fig.~\ref{figSI:cont}, we employ a semi-implicit time-stepping scheme: the stiff fourth-order diffusion term is treated implicitly in Fourier space, while the nonlinear and advective terms are evaluated explicitly in real space.
Periodic boundary conditions are imposed on a square domain of size $\tilde{L}=25.6$, with grid spacing $\Delta \tilde{x}=0.05$ and time step $\Delta \tilde{t}=0.0025$. The parameters are set to $\tilde{\gamma}=1$, $\tilde{\zeta}=0.01$, and $\tilde{\kappa}=0.025$, for which the characteristic interfacial thickness is 
$2\sqrt{2\tilde{\kappa}}=0.45$.
The initial conditions are
\begin{align}
\tilde{\rho} &= 0.5 + \xi_\varepsilon, \\
\bm{p} &= \bm{0},
\end{align}
where $\xi_\varepsilon$ is a uniformly distributed random number in 
$[-0.05,\,0.05]$.
This choice ensures that the homogeneous state is weakly perturbed,
allowing spontaneous pattern formation to develop.

The numerical results are shown in Fig.~\ref{figSI:cont} 
for $\tilde{w}=0$ (a), $\tilde{w}=0.001$ (b) and $\tilde{w}=-0.001$ (c). For $\tilde{w}=0$, the model reduces to the standard Cahn--Hilliard equation, corresponding to the monodisperse system. Aggregates coarsen over time and eventually become stationary, as evidenced by the vanishing temporal difference of the density field $\delta \tilde{\rho}$.

We found almost identical coarsening behavior when $\tilde{w}=-0.001$. In contrast, for $\tilde{w}=0.001$, aggregates initially undergo fragmentation. Although larger clusters subsequently re-form, the background density remains finite, indicating the persistent presence of small clusters, as also observed in the particle-based simulations. Moreover, the aggregates do not settle into a static configuration but continue to move, as seen from the nonzero temporal difference fields $\delta \tilde{\rho}$.

This interfacial mechanism explains the numerical observations. Since the density inside aggregates is nearly homogeneous due to Cahn--Hilliard-type segregation, polarity remains localized at interfaces. For $\tilde{w}>0$, the resulting outward polarity destabilizes aggregates and promotes fragmentation, whereas for $\tilde{w}<0$ it suppresses fragmentation and stabilizes clusters. 
These two regimes correspond to the head--large and tail--large geometries observed in the particle-based model and experiments.

To rationalize these observations, we consider the unperturbed reference state corresponding to $\tilde{w}=0$.
}
\fixred{For a flat interface (1D profile), it gives the standard tanh profile
\begin{align}
\tilde{\rho}(\tilde{x})=\frac{1}{2}\left[1-\tanh\!\left(\frac{\tilde{x}-\tilde{x}_0}{2\sqrt{2\tilde{\kappa}}}\right)\right].
\label{eq:tanh_profile}
\end{align}

A radially symmetric droplet in two dimensions can be constructed in the large-radius limit $\tilde{R}\gg \sqrt{\tilde{\kappa}}$ 
by replacing the coordinate normal to the interface with the signed distance from a circle,
$\tilde{x}-\tilde{x}_0 \rightarrow \tilde{r}-\tilde{R}$. 
This yields the approximate droplet profile
\begin{align}
\tilde{\rho}(\tilde{r})\approx \frac{1}{2}\left[1-\tanh\!\left(\frac{\tilde{r}-\tilde{R}}{2\sqrt{2\tilde{\kappa}}}\right)\right],
\label{eq:droplet_profile}
\end{align}
which is accurate up to curvature corrections of order $\mathcal{O}(\sqrt{\tilde{\kappa}}/\tilde{R})$.

Using Eq.~\eqref{eq:droplet_profile}, the polarity dynamics becomes
\begin{align}
\partial_{\tilde{t}} \bm{p}
= -\frac{\tilde{\zeta}}{\tilde{\gamma}}\bm{p}-\frac{\tilde{w}}{\tilde{\gamma}}\tilde{\nabla}\tilde{\rho}.
\end{align}
Assuming fast relaxation of the polarity field, we obtain
\begin{align}
    \bm p \simeq -\frac{\tilde w}{\tilde\zeta}\tilde\nabla\tilde\rho,
\end{align}
which implies that $\bm{p}$ is generated predominantly in the interfacial region 
$\tilde{r}\sim \tilde{R}$, where $\tilde{\nabla}\tilde{\rho}$ is finite. 
The polarity points outward for $\tilde{w}>0$ 
and inward for $\tilde{w}<0$, 
thereby destabilizing or stabilizing the droplet, respectively.


Thus, the continuum model reproduces the essential qualitative features observed in experiments and agent-based simulations. In particular, positive $\tilde{w}$ induces outward interfacial polarity that destabilizes aggregates, leading to fragmentation and persistent motion, whereas negative $\tilde{w}$ stabilizes clusters. These results identify polarity generated at aggregate interfaces as the minimal mechanism that controls the stability and motility of clusters in the nonreciprocal colloidal system.

We emphasize that the present formulation is valid only for sufficiently small $|\tilde{w}|$. For larger values, the model develops short-wavelength instabilities that are likely artifacts of the truncated gradient expansion. Regularization by higher-order gradient terms or an extended two-density formulation would be required for a more complete theory, which we leave for future work.
}

\fixred{
\section{Extension to the {\it dimer} model}\label{SI:dimer}
To describe permanently bonded dimers as reported in Refs.~\cite{Ma2015, Niu2017, Niu2018, Wang2014-is, Shields2017-zv, Diwakar2022-qp, Diwakar2024-go}, we extend the agent-based model defined by 
Eqs.~\eqref{eq:ndimEq}-\eqref{Eq:ndimFEHD}.

We prepare $N$ dimers composed of particles I and II, for a total of $2N$ particles with $N_\mathrm{I}=N_\mathrm{II}=N$. Particles with indices $i\in[0,N-1]$ correspond to species I, and $i\in[N,2N-1]$ to species II. Particles $i$ and $i+N$ form a permanently bonded dimer.
}

\begin{figure}
    \centering  
    \includegraphics{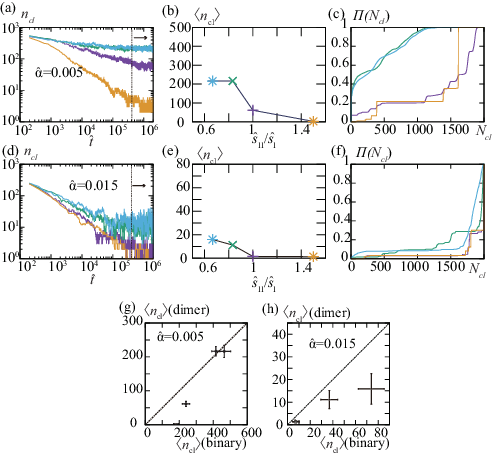}
    \caption{\fixred{
    Cluster-size analysis of the dimer model in a domain $\hat{L}_x=\hat{L}_y=24$ with $N_\mathrm{I}=N_\mathrm{II}=1000$. Results for $\hat{\alpha}=0.005$ (a-c) and $\hat{\alpha}=0.015$ (d-f) are shown. EHD interaction parameter were fixed at $(\hat{\ell}_\mathrm{I},\hat{\ell}_\mathrm{II})=(1/6,\,1/9)$. (a,d) Time evolution of the number of clusters $n_\mathrm{cl}$. (b,e) Time-averaged cluster number $\langle n_\mathrm{cl}\rangle$. (c,f) Cumulative cluster-size distribution $\Pi(N_{cl})$. 
    (g,h) Comparison of $\langle n_\mathrm{cl}\rangle$ between the dimer and bidisperse systems for $\hat{\alpha}=0.005$ (g) and $0.015$ (h). Error bars denote temporal standard deviation.
    }
    }
    \label{figSI:dimerCluster}
\end{figure}

\fixred{
The dynamics are modified only in the steric interaction term. Equation~\eqref{eq:ndimFcol} is replaced by
\begin{align}
\hat{\bm{F}}_{ij}^{\mathrm{col}}
=(\hat{s}_i+\hat{s}_j-|\hat{\bm{r}}_{ij}|)\,
\dfrac{\hat{\bm{r}}_{ij}}{|\hat{\bm{r}}_{ij}|},
\label{eq:Fcoldim1}
\end{align}
when $(i\equiv j \ \mathrm{mod}\ N)$, which imitate permanent bonding between paired particles, and
\begin{align}
\hat{\bm{F}}_{ij}^{\mathrm{col}}
=
\begin{cases}
(\hat{s}_i+\hat{s}_j-|\hat{\bm{r}}_{ij}|)\,
\dfrac{\hat{\bm{r}}_{ij}}{|\hat{\bm{r}}_{ij}|},
& (|\hat{\bm{r}}_{ij}|\le \hat{s}_i+\hat{s}_j), \\
0,
& (|\hat{\bm{r}}_{ij}|>\hat{s}_i+\hat{s}_j),
\end{cases}
\label{eq:Fcoldim2}
\end{align}
for all other particle pairs. All other model parameters are identical to those in Eqs.~\eqref{eq:ndimEq}-\eqref{Eq:ndimFEHD}.

To compare with the bidisperse system analyzed in Sec.~\ref{SI:ClusterSize}, we fixed $(\hat{\ell}_\mathrm{I},\hat{\ell}_\mathrm{II})=(1/6,\,1/9)$ which controls the interaction through EHD flow, while varying the steric radii $(\hat{s}_\mathrm{I},\hat{s}_\mathrm{II})$. Simulations were performed with $N_\mathrm{I}=N_\mathrm{II}=1000$ ($2000$ particles in total), for $\hat{\alpha}=0.005$ and $0.015$.

The results are summarized in Fig.~\ref{figSI:dimerCluster}. The temporal evolution of the cluster number $n_\mathrm{cl}$ shows saturation at long times, and the time-averaged value $\langle n_\mathrm{cl}\rangle$ decreases as the steric asymmetry increases.

Compared with the bidisperse system, 
the dimer model exhibits a systematic reduction of $\langle n_\mathrm{cl}\rangle$, indicating enhanced aggregation. For sufficiently large $\hat{s}_\mathrm{II}/\hat{s}_\mathrm{I}$, near-complete aggregation is observed.

This behavior is consistent with enhanced motility-induced clustering, as reported in related active dimer systems~\cite{Chaki2026-mz}. A more detailed characterization of the cluster statistics is left for future investigation.
}

\end{document}